%% file: manuscript.tex
\RequirePackage{fix-cm}
\documentclass[smallextended]{svjour3}\smartqed\usepackage{graphicx}
\usepackage{framed}
\usepackage{enumitem}
\usepackage{tabularx}

\usepackage{booktabs}

\usepackage[table, dvipsnames]{xcolor}

\usepackage{float}
\usepackage{subcaption}

\usepackage[colorlinks]{hyperref}
\hypersetup{colorlinks,breaklinks,
            urlcolor=[rgb]{0.85098,  0.25490,  0.12157},
            linkcolor=[rgb]{0.12549,  0.33333,  0.85882},            citecolor=blue}

\usepackage[style=numeric, backend=biber, defernumbers=true, maxbibnames=99, giveninits=true]{biblatex}
\def\makeheadbox{{\hbox to0pt{\vbox{\baselineskip=10.7dd\hrule\hbox
to\hsize{\vrule\kern3pt\vbox{\kern3pt
\hbox{This is a \href{https://creativecommons.org/licenses/by/4.0/}{CC-BY} licensed version of the work:}
\hbox{Winkler, D., Urbanke, P. {\&} Ramler, R. Investigating the readability of test}
\hbox{code. \textit{Empir Software Eng} \textbf{29}, 53 (2024). \href{https://doi.org/10.1007/s10664-023-10390-z}{DOI:10.1007/s10664-023-10390-z}}
\kern3pt}\hfil\kern3pt\vrule}\hrule}\hss}}}

\usepackage[normalem]{ulem}
\definecolor{darkBlue}{RGB}{69,92,221}

\begin{document}

\title{Investigating the Readability of Test Code}

\subtitle{Combining Scientific and Practical Views}

\author{Dietmar Winkler         \and
        Pirmin Urbanke          \and
        Rudolf Ramler
}

\institute{Dietmar Winkler\at
              SBA Research gGmbH, Floragasse 7, Vienna, 1040, Austria; \\
              Austrian Center for Digital Production, Seestadtstrasse 27/19, Vienna, 1200, Austria; and \\  
              Institute for Information Systems Engineering, TU Wien, Favoritenstr. 9/194, Vienna, 1040, Austria \\
              https://orcid.org/0000-0002-4743-3124 \\
              \email{dietmar.winkler@tuwien.ac.at}           \and
            Pirmin Urbanke \at
              Christian Doppler Laboratory on Security and Quality Improvement in the Production Systems Lifecycle (CDL-SQI), TU Wien, Favoritenstrasse 9/194, Vienna, 1040, Austria \\
              https://orcid.org/0009-0005-1868-3078\\
              \email{pirmin.urbanke@tuwien.ac.at}                
           \and
           Rudolf Ramler \at
              Software Competence Center Hagenberg GmbH (SCCH), 
              Softwarepark 32a, 4232 Hagenberg, Austria \\
              https://orcid.org/0000-0001-9903-6107\\
              \email{rudolf.ramler@scch.at}
}

\date{Accepted: 30 August 2023, Published: 26 February 2024}

\maketitle

\begin{abstract}

\textbf{Purpose.}
The readability of source code is key for understanding and maintaining software systems and tests.  
Although several studies investigate the readability of source code, there is limited research specifically on the readability of test code and related influence factors.
\textbf{Objective.} In this paper, we aim at investigating the factors that influence the readability of test code from an academic perspective based on scientific literature sources and complemented by practical views, as discussed in grey literature.   
\textbf{Methods.} First, we perform a Systematic Mapping Study (SMS) with a focus on scientific literature. Second, we extend this study by reviewing grey literature sources for
practical aspects on test code readability and understandability. 
Finally, we conduct a controlled experiment on the readability of a selected set of test cases to collect additional knowledge on influence factors discussed in practice.
\textbf{Results.}
The result set of the SMS includes 19 primary studies from the scientific literature for further analysis.
The grey literature search reveals 62 sources for information on test code readability.
Based on an analysis of these sources, we identified a combined set of 14 factors that influence the readability of test code. 7 of these factors were found in scientific \textit{and} grey literature, while some factors were mainly discussed in academia (2) \textit{or} industry (5) with only limited overlap. 
The controlled experiment on practically relevant influence factors showed that the investigated factors have a significant impact on  readability for half of the selected test cases. \textbf{Conclusion.} Our review of scientific and grey literature showed that test code readability is of interest for academia and industry with a consensus on key influence factors.  
However, we also found factors only discussed by practitioners. For some of these factors we were able to confirm an impact on readability in a first experiment. 
Therefore, we see the need to bring together academic and industry viewpoints to achieve a common view on the readability of software test code.

\end{abstract}

\section{Introduction}
\label{intro}
\textit{Software Tests} encode important knowledge about typical usage scenarios and inputs, corner cases, exceptional situations as well as the intended output and behavior of the software system. Consequently, test cases play an important role for assuring software quality and also for the evolution of software systems as a knowledge source that supports communication in the team and with customers~\cite{latorre2014successful} and for specification and documentation~\cite{ricca2009using}. At the same time, the evolution of software systems requires that test cases are frequently updated and extended, resulting in effort for corresponding test case evolution and maintenance activities~\cite{moonen2008interplay,zaidman2011studying}. 
\textit{Test Code}~\cite{garousi2016developing}, i.e., the form in which executable automated tests are commonly available, is the basis for many downstream activities such as maintaining and refactoring tests, locating faults, debugging, analyzing and comprehending test results, repairing broken tests, or dealing with flakiness~\cite{garousi2016developing}.
In all these scenarios, developers and testers have to repeatedly read and understand test code -- a usually time-consuming manual task, which makes readability and understandability critical factors when it comes to the quality of a project's test cases~\cite{kochhar2019practitioners} \cite{Setiani2021251NormalBib}.

\textit{Readability} as well as legibility and understandability of source code have already been subject to a series of empirical studies, which were recently examined in a systematic literature review by Oliveira \textit{et al.}~\cite{oliveira2020evaluating}. Test code has many properties in common with source code of software programs, and tests are often written using the same programming languages as the system under test. Nevertheless, the development of test code also shows significant differences when compared to other code. There exist dedicated frameworks and patterns for implementing test code~\cite{meszaros2007xunit} and, furthermore, tools for generating tests (e.g., Evosuite, Randoop, IntelliTest) are becoming increasingly popular~\cite{ramler2018applying}.

In this paper, we focus on \textbf{investigating the readability of software test code by combining scientific and practical views}.
In a first step, we build on a Systematic Mapping Study (SMS) approach~\cite{petersen2015guidelines} to identify characteristics, influence factors, and assessment criteria that have an impact on the readability of test code. 
In a second step, we complement the mapping study with grey literature to include practical views. 
Based on identified influence factors, we conducted a controlled experiment~\cite{wohlin2012experimentation} to investigate the perception of readability and understandability in academic environment.

In an initial mapping study~\cite{winkler2021vst} we reviewed the scientific literature dedicated to the readability of test code, exploring (a) the demographics of the body of knowledge, (b) the characteristics of the studied test code, and (c) the factors that have shown to impact readability. We analyzed 19 scientific studies filtered from several hundred search results and identified a set of 9 influence factors that have been investigated in academic work either individually or as part of comprehensive readability models.
Our mapping study covers the topic of test code readability specifically from the viewpoint of work published in the scientific literature. However, test code readability is of high practical relevance and the topic is therefore also frequently covered in magazine articles, books on testing, and online blogs. These sources are typically referred to as grey literature~\cite{garousi2019guidelines}. 
Based on previous work~\cite{winkler2021vst}, the goal of this work is to extend the topic by  exploring  test code readability in it's entire scope by combining the scientific and the practical viewpoint. We therefore conducted an additional grey literature survey to identify influence factors commonly discussed in practice, we mapped the results to the findings from the previous scientific literature study. 
Finally, we investigated the newly identified factors in a controlled experiment~\cite{wohlin2012experimentation} comparing the readability of different versions of a selected set of test cases.
The main contribution of the paper includes:
\begin{enumerate}
    \item \textbf{Identified influence factors} for the readability of software test code based on a systematic mapping study (SMS) as a combination of academic and practical views, derived from academic and grey literature. Detailed analysis results are available online~\cite{dataset2023}\footnote{\label{footnote1}Data Set~\cite{dataset2023}: ~\url{https://doi.org/10.48436/w4q8v-28695}
    }.
    \item \textbf{Setup and results of a controlled experiment} to investigate influence factors of a selected set of test cases in an academic environment~\cite{dataset2023}.
    
\end{enumerate}

Consequently, the remainder of this paper is structured as follows: \hyperref[rw]{Section~\ref*{rw}} describes background and related work on test code quality and code readability. 
\hyperref[rq]{Section~\ref*{rq}} defines our research questions, followed by three sections explaining the setup, process and results of the systematic mapping study (\hyperref[rq]{Section~\ref*{sec:sms}}), the grey literature survey (\hyperref[rq]{Section~\ref*{sec:grey-literature}}), and the concluding experiment (\hyperref[rq]{Section~\ref*{sec:experiment}}) in context of the respective research questions. 
Finally, \hyperref[rq]{Section~\ref*{discussion}} summarizes the finding, discusses implications for academia and practitioners and limitations, and identifies future work.

\section{Background}
\label{rw}

The readability of test code is associated to two areas of related research: First, the area of test quality or, more specifically, the quality of code of automated tests (cf. Section~\ref{sec:21}). Second, the related research on source code readability (cf. Section~\ref{sec:22}).

\subsection{Software Test Code Quality}
\label{sec:21}
In context of software evolution and maintenance, changes made to the software due to bug fixes, extensions, enhancements, and improvements, also require subsequent adaptations of the corresponding tests~\cite{yusifouglu2015software}. 
Thereby, similar to code quality being an important factor for supporting evolution and maintenance, test code quality is critical for evolving and maintaining tests. 
Consequently, in test code engineering~\cite{yusifouglu2015software}, the two leading activities are quality assessment and co-maintenance of test-code with production code. 

Engineering test code, much like engineering production code, is a challenging process and prone to all kinds of design and coding errors. Hence, test code also contains bugs, which may either cause false alarms (i.e., a test fails although the production code is correct) or which may cause ``silent horrors" (i.e., a test passes although the production code is incorrect). Both kind of bugs have been found to be prevalent in practice~\cite{vahabzadeh2015empirical}. The latter kind of bugs are also considered a result of ``rotten green tests"~\cite{delplanque2019rotten}, which are tests that pass green but do so by inadequately validating the required properties of the system under test. 

Apart from bugs in tests, a widely reported problem related to test code quality are \textit{test smells}~\cite{garousi2018smells,spadini2018relation,tufano2016empirical}. Test smells are the equivalent to code smells~\cite{lacerda2020code} (or anti-patterns) in production code, which are symptoms of an underlying problem in the code (e.g., a design problem) that may not cause the software to fail now but bears the risk of causing additional problems and actual bugs in future. Hence, test smells can be considered as poorly-designed tests (similar to rotten green tests) and their presence may negatively affect test suites with respect to the  maintainability and even the correctness of the tests ~\cite{bavota2015test,spadini2018relation}. Although test smells are a popular concept that is frequently investigated in scientific literature, a recent study by Panichella \textit{et al.} \cite{panichella2022test} suggests a mismatch between the definition of test smells and real problems in the tests. To tackle this mismatch they update definitions of test smells and investigate issues which are currently not covered well by the existing smells.

Since the upcoming of test code generators like Evosuite, which aim to generate test suites covering the whole system under test, there is a continuous discussion on improvements and practical usefulness of these tools. For example McMinn \textit{et al.} \cite{McMinn2012SearchBased} leverage web search engines for generating test data of type \texttt{String}. This approach can improve the coverage of the resulting test suites and the use of more realistic input strings could improve the readability of the test code. Various studies \cite{fraser2013whiteboxtest,ceccato2015autogentest,shamshiri2018autogentest} investigate the usefulness of test code generators in debugging activities and also highlight shortcomings of generated test code which relate to the high number of assertions, absence of explanatory comments or documentation, quality of identifiers and in general unrealistic test scenarios. Hence, similar to the quality of automatically generated code~\cite{yetistiren2022assessing,al2022readable}, the quality of generated test code is a critical aspect that requires additional consideration and investigation.

\subsection{Readability of Source Code}
\label{sec:22}

Reading and understanding source code is an elementary activity in software maintenance and development~\cite{minelli2015know}. Hence, code readability has been subject to a wealth of empirical studies; e.g., Oliveira \textit{et al.}~\cite{oliveira2020evaluating} examining 54 papers on code readability in their literature review. In these studies, a wide range of different factors influencing readability have been investigated, including code formatting and indentation, identifier naming, line length, complexity of expressions, complexity of the control flow, use of code comments, presence of code smells, and many others. 

Buse \textit{et al.}~\cite{buse2008metric} developed a model combining aforementioned factors to automatically estimate code readability. They trained their model on small source code snippets extracted from open-source projects and tagged as readable or non-readable by human annotators. Following this approach, generalized and extended code readability models have been developed in subsequent works, e.g., by Posnett \textit{et al.}~\cite{posnett2011simpler} and Scalabrino \textit{et al.}~\cite{scalabrino2017automatically}.

Buse \textit{et al.} define \textit{readability} as \textit{``a human judgment of how easy a text is to understand"}~\cite{buse2008metric}. However, no generally accepted definition for the term readability has been established in the literature and, thus, readability is often used in combination with or as synonym for the related terms \textit{legibility} and \textit{understandability}. The term legibility is rather related to the visual appearance of the source code affecting the ability to identify individual elements~\cite{oliveira2020evaluating}, while the term understandability is mainly related to semantic aspects of the code. Scalabrino \textit{et al.}~\cite{scalabrino2017automatically} even developed a model specifically dedicated to the understandability of source code, arguing that program understanding is a non-trivial mental process that requires building high-level abstractions from code to understand its purpose, relationships between code entities, and the latent semantics, which all cannot be sufficiently captured by readability metrics alone. 

 \begin{figure}[ht]
    \centering
    \includegraphics[scale=0.95]{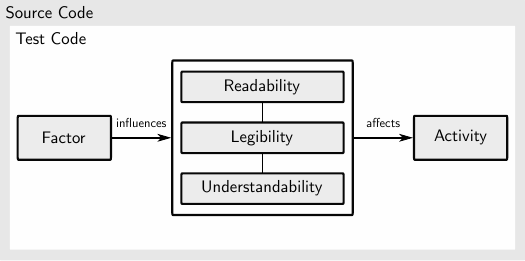}
    \caption{Concept of readability as used in context of our study.}
    \label{fig:concept}
\end{figure}

Based on the terms and definitions commonly used in  related work, we adopt a broader view on the concept of readability embracing all three terms -- \textit{readability}, \textit{legibility} and \textit{understandability} -- in our work in context of test code readability. Hence, in the remainder of this paper, we use the term readability subsuming all related notations since it is the most commonly used term in the software engineering literature.

\autoref{fig:concept} depicts this view and the distinction between factors (e.g., test case length) influencing readability as well as upstream activities (e.g., test case maintenance) being affected by readability. The underlying concept is related to activity-based quality modeling as proposed in Quamoco approach~\cite{wagner2015operationalised,wagner2012quamoco} and used, e.g., for modeling maintainability~\cite{deissenboeck2007activity} or requirements quality~\cite{femmer2015s}. Readability is described by more fine-grained factors that can be assessed (manually or automatically) and it has an observable impact on the activities performed by stakeholders related to a specific entity. In our context, the entity of interest is the test code, shown as a subset of source code.

\section{Research Questions}
\label{rq}
Based on the goal of this article to investigate the readability of software test code by combining scientific and practical views, we identified three groups of research questions, with focus on (a) influence factors in academia, (b) influence factors in practice, and (c) an investigation of combined influence factors on a selected set of test cases in a controlled experiment.

\subsection{Influence Factors in Academia}
Based on our previous work, an initial Systematic Mapping Study (SMS)~\cite{winkler2021vst}, we extended the mapping study by introducing additional analysis criteria. Therefore, we identified the first research question (RQ1) and two sub-research questions to (a) identify influence factors (RQ1.1) and (b) to explore methods (RQ1.2) used in scientific studies.
We applied the Systematic Mapping Study (SMS) approach, proposed by Petersen \textit{et al.}~\cite{petersen2015guidelines}. Section~\ref{sec:sms} describes the research protocol and the results of the mapping study. 

\begin{framed}
    \noindent \textit{RQ1. Influence factors on test code readability in scientific literature?}
    
    \textit{RQ1.1 Which influence factors are analyzed in scientific literature?}
    
    \textit{RQ1.2 Which methods are used in scientific studies?}
\end{framed}

\subsection{Influence Factors in Practice.} To systematically capture influence factors, discussed in industry and practice, we extended and complemented the mapping studies with grey literature.
The results will show similarities and differences of academia and industry in context of the readability of software test code. 
For investigating grey literature, we followed the guidelines proposed by Garousi~\textit{et al.}~\cite{garousi2019guidelines}. 
Section~\ref{sec:grey-literature} presents the research protocol and the results.

\begin{framed}
    \noindent \textit{RQ2. Influence factors on test code readability discussed in practice?}
    
    \begin{description}
    \item \textit{RQ2.1 Which influence factors are discussed in grey literature?}
    
    \item \textit{RQ2.2 What is the difference between influence factors in scientific literature and grey literature?}
    \end{description}
\end{framed}

\subsection{Investigating Influence Factors in a Controlled Experiment.} 
Based on identified factors, we conducted a controlled experiment in academic environment to investigate the perception of readability and understandability of a selected set of test cases (derived from open source projects). 
We build on the guidelines, proposed by Wohlin ~\textit{et al.}~\cite{wohlin2012experimentation} for planning, executing, and reporting on the controlled experiment. Section~\ref{sec:experiment} presents the experiment setup and reports on the results.
\begin{framed}
    \noindent \textit{RQ3. What is the influence on the test code readability of discussed factors?}
    \begin{description}
    \item \textit{RQ3.1 Do factors discussed in practice show an influence on readability when scientific methods are used?}
    \end{description}
\end{framed}

\section{Systematic Mapping Study}
\label{sec:sms}
To investigate the \textit{Influence Factors on Readability in Scientific Literature}, we conducted a \textit{Systematic Mapping Study (SMS)} based on Petersen ~\textit{et al.}~\cite{petersen2015guidelines}. 
In this section, we summarize the study protocol and the results
 from the systematic mapping study (SMS)~\cite{dataset2023}. We present influence factors and  study types with focus on research methods used.

\subsection{Study Protocol and Process}
This section summarizes the study protocol with focus on the systematic mapping study of scientific publications.

\begin{figure}[ht]
    \centering
    \includegraphics[scale=0.84]{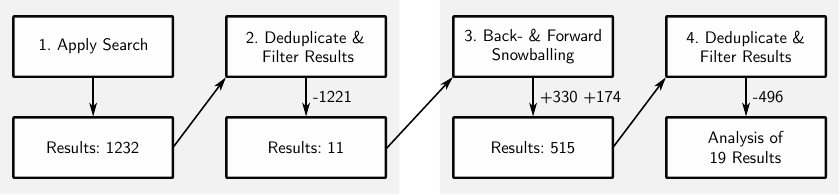}
    \caption{Systematic mapping study process and amount of received publications.}
    \label{fig:sms-process}
\end{figure}
An integral part of systematic mapping studies as proposed by Petersen \textit{et al.} \cite{petersen2015guidelines} is the thorough documentation of the process to make the results traceable. \autoref{fig:sms-process} provides an overview of the whole process. After the initial search (Step~1) and filtering (Step~2), we apply back- and forward snowballing (Step~3) and filter (Step~4) to obtain our final set of studies. We repeat the steps 3 and 4 , the back- and forward snowballing, until exhaustion, i.e., they add no new relevant studies to the result set. In our case, no additional publications were identified in the second iteration. The following subsections provide details on each of these steps.

\paragraph{\textbf{Step 1: Apply Search.}}
Based on the research questions (cf. Section~\ref{rq}, RQ1), we defined the following \textit{keywords}: \textit{test, code, model, readability, understandability, legibility} and \textit{smell}. We used them to build the \textit{Search Strings} shown in \autoref{tab:search-string}.
The queries were performed on established sources for scientific literature, i.e., \textit{Scopus}, \textit{IEEE} and \textit{ACM}, and we filtered the studies based on title, abstract and keywords.
In the ACM search we enclosed the term ``understandability" in double quotes in the abstract filter to enforce exact matching, because ACM's fuzzy matching leads to a high number of irrelevant results.
For ACM we searched in the \textit{ACM Guide to Computing Literature} which offers a larger search space than the \textit{ACM Full-Text Collection}. 
The search was conducted at the end of November 2021 without limiting the publication year and returned a total of 1232 raw results (Scopus: 460, IEEE: 231, ACM:541). Based on the merged results, we proceeded to the next step.

\begin{table}[t]
    \centering
    \caption{Search strings in different databases.}
    \input{search_strings.tex}
    \label{tab:search-string}
\end{table}

\paragraph{\textbf{Step 2: Deduplicate \& Filter Results.}}

We first deduplicated the raw results based on the digital object identifier (DOI) and title, which removed 281 studies. Next, we imported the result set into a spreadsheet solution for applying inclusion and exclusion criteria. \\

\noindent \textbf{Inclusion Criteria.} We included a study if both of the following criteria were fulfilled:

\begin{itemize}
    \item Conference papers, journal/magazine articles, or PhD theses (returned by ACM)
    \item Readability, understandability or legibility of test code is an object of the study
\end{itemize}

\noindent\textbf{Exclusion Criteria.} We excluded a study if one of the following criteria is applicable:

\begin{itemize}
    \item Not written in English
    \item Conference summaries, talks, books, master thesis
    \item Duplicate or superseded studies    \item Studies not identifying factors that influence test code readability
\end{itemize}

The criteria were evaluated based on title and abstract of the results by at least one of the authors. When in doubt about including or excluding, the evaluated study was discussed with a second evaluator. This step left us with 11 scientific publications.

\paragraph{\textbf{Backward \& Forward Snowballing - First Iteration. }} Based on the initial iteration, we executed backward \& forward snowballing to identify relevant studies that have not been identified in the initial search.

 \begin{itemize}
     \item \textbf{Step 3: Backward \& Forward Snowballing.} Since relevant literature might refer to further important studies, we used the references included in the 11 studies for backward snowballing via Scopus. The 11 studies might also be cited by other relevant studies, hence we also performed forward snowballing, by using Scopus to find studies, which cite one of the initial 11 studies. This increased the result set by 330 from backward snowballing and 174 from forward snowballing to a total of 515 studies.

     \item \textbf{Step 4: Deduplicate \& Filter Results.} By comparing these 515 studies with the initial result set we found and removed 83 duplicates. Similar to step 2, one of the authors of this paper applied the inclusion and exclusion criteria. Additionally, after a full text reading, all studies were discussed and reevaluated by the author team. With this, we reduced the result set by 496 and obtained a final result of 19 studies.
 \end{itemize}

\paragraph{\textbf{Backward \& Forward Snowballing - Second Iteration. }}
We performed a second iteration of back- and forward snowballing via Scopus with these 19 studies as input. This returned a raw result of 825, which we reduced to 357 studies by removing duplicates. We applied in- and exclusion criteria on the remaining studies, which removed all 357 studies. Therefore, this second iteration did not add any new relevant studies to the results set of 19 studies.

\paragraph{\textit{Studies Not Included.}}
In the following, we provide four exemplary cases filtered out in step 2 and the rationale why these studies did not meet the criteria for inclusion in the final publication set after discussion by all authors:
Grano \textit{et al.}\cite{grano2020pizza} focus on semi-structured interviews with five developers from industry and a confirmatory online survey to synthesize which factors matter for test code quality. Although readability is seen as a critical factor by all participants, the analysis of readability and influencing factors was not in the scope of this work.
Tran \textit{et al.} \cite{tran2019test} investigated general factors for test quality by interviewing 6 developers from a company. Quality factors are discussed with natural language tests brought by the participants. Since our work has its specific focus on test code, readability of natural language tests was not considered further.
Bavota \textit{et al.} \cite{bavota2015test} report on four lab experiments with an overall number of 49 students and 12 practitioners and effects on maintenance tasks from eight test smells. These test smells occur frequently in software systems. While this work clearly shows that test smells negatively affect correctness and effort for specific maintenance tasks, a connection between test smells and readability is not shown.
Dei\ss~\cite{deiss2008refactoring} reports on a case study about semi-automatic conversion and refactoring of a TTCN-2 test suite to TTCN-3. Discussed improvements included reducing complicated or unnecessary code artifacts generated by the automatic conversions that are also supposed to increase readability. We excluded this study since the focus was the migration from TTCN-2 to TTCN-3 and the study did not investigate the improvements on the readability of test code.

\subsection{Systematic Mapping Study Results}
This section summarizes the findings in context of influence factors on readability found in scientific literature.

\begin{table}[ht]
    \centering
    \caption{Final Set of Publications based on the Search Process.}
    \input{selected_papers.tex}
    \label{tab:selected-studies}
\end{table}

\subsubsection{Which influence factors are analyzed in scientific literature (RQ1.1)?}
\label{subsub:RQ1.1-results}

In RQ1, we explore the factors that have been found to impact readability of test code. 
Influence factors have been derived by one of the authors based on the content of the paper. Other authors and testing experts have reviewed the initially identified factors. Deviations have been discussed by all authors to come to a consensus.
\autoref{tab:factors} maps candidate factors to the studies that investigate them. Two approaches of how influence factors are considered in the primary studies can be distinguished. Studies either \textbf{(a)} investigate the impact of one or more \textit{individual factors}, often related to the attempt to improve readability with a specific approach or tool, or \textbf{(b)} they target \textit{readability models} constructed from a combination of many factors. The majority of the primary studies (see \autoref{tab:factors-subtab-a}) consider individual factors. Readability models were subject to study only in three instances (\autoref{tab:factors-subtab-b}), although such models are commonly used in the general research on source code readability.

\begin{table}[H]
    \centering
    \caption{Reported factors influencing test code readability.}
    \begin{subtable}{\linewidth}
        \caption{Studies investigating individual factors.}
        \input{factors.tex}
        \label{tab:factors-subtab-a}
    \end{subtable}
   \vspace*{\floatsep}
    \begin{subtable}{\linewidth}
        \caption{Studies using readability models.}
        \input{factors_models.tex}
        \label{tab:factors-subtab-b}
    \end{subtable}
    \label{tab:factors}
\end{table}

We identified a total of 9 unique influence factors in the scientific literature as shown in \autoref{tab:factors-subtab-a} and \autoref{tab:factors-subtab-b}. In the following, we briefly explain these factors. The number in brackets shows the number of primary studies including the particular factor combining counts from both tables.

\begin{description}
    \item \textbf{Test names (6)}: The name of the test method or test case. Not only generated tests have poor names but also names provided by humans often convey few useful information. Therefore, several studies propose different solutions on automatic test renaming e.g., Zhang \textit{et al.} \cite{Zhang2016625} Roy \textit{et al.} \cite{Roy2020287} or Daka \textit{et al.} \cite{Daka201757}. In studies from e.g. Setiani \textit{et al.} \cite{Setiani2021251}, Bowes \textit{et al.} \cite{Bowes20179} or Panichella \textit{et al.} \cite{Panichella2016547}, participants agree on the importance of test names for readability.
    \item \textbf{Assertions (5)}: This factor relates to the amount of assertions in a test case as well as to assertion messages. Although assertions are an integral part of test code Daka \textit{et al.} \cite{Daka2015107} report minor influence on readability coming from the amount of assertions in a test. Nevertheless the amount of assertions is still used in their readability model and also in the model from Setiani \textit{et al.} \cite{Setiani2020169036}. In the survey from Setiani \textit{et al.} \cite{Setiani2021251} assertions are mentioned to have an influence, but other factors like naming are deemed more important. In Almasi \textit{et al.} \cite{Almasi2017263} developers issued concerns on generated assertions. Leaotta et al. \cite{Leotta2018AssertJ} find no significant effects on readability when AssertJ assertions are used instead of basic JUnit assertions, although other positive effects could be observed.

    \item \textbf{Identifier names (5)}: Naming of variable names in test cases. Especially Lin \textit{et al.} \cite{Lin2019204} investigate this factor thoroughly and also provide characteristics of good and bad identifier names based on a survey. Roy \textit{et al.} \cite{Roy2020287} propose an automatic way for identifier renaming in test cases. In studies from e.g. Setiani \textit{et al.} \cite{Setiani2021251} and Bowes \textit{et al.} \cite{Bowes20179} participants agree on the importance of identifier names for readability. Fisher and Johnson \cite{Fisher2018800} attribute differences between generated and manually written tests to differences in naming.
    
    \item \textbf{Test structure (2+3)}: Structural features are found in studies investigating individual factors (2 times) as well as in readability models (3 times). They include strucural features of test methods like maximum line length, number of identifiers, length of identifiers, number of control structures (e.g., branching as mentioned in Bowes \textit{et al.} \cite{Bowes20179}), test length, etc. Participants in the study from Setiani \textit{et al.} \cite{Setiani2021251} agree on the importance of the amount of lines of code in the tests. These features are also used in combination by automatic readability raters, e.g., from Daka \textit{et al.} \cite{Daka2015107}, who propose a rater especially for test cases, Grano \textit{et al.} \cite{Grano2018348} or Setiani \textit{et al.} \cite{Setiani2020169036}.

    \item \textbf{Test data (4)}: Testers often have to evaluate data used in assertions to decide if a test has truly failed or if there is a fault in the test. Afshan \textit{et al.} \cite{Afshan2013352} investigates this topic and shows that readable string test data helps humans predicting correct outcome. Alsharif \textit{et al.} \cite{Alsharif2019SQL} and Almasi \textit{et al.} \cite{Almasi2017263} highlight the importance of meaningful test data. Furthermore, in the workshop study from Bowes \textit{et al.} \cite{Bowes20179} developers, amongst others, state that \textit{magic numbers} are detrimental to readability.

    \item \textbf{Test summaries (4)}: Documentation describing the whole test case support understanding what the tests do, for example as Javadoc like in Roy \textit{et al.} \cite{Roy2020287} or Li \textit{et al.} \cite{Li2016341} or interleaved with test code like in Panichella \textit{et al.} \cite{Panichella2016547}. Li \textit{et al.} \cite{Li2018Aiding} reduce the amount of generated description, by only adding test stereotypes as tags.

    \item \textbf{Dependencies (3)}: The number of classes a single test case depends on, as proposed by Fraser \textit{et al.} \cite{Fraser201180}, or if a test is truly a unit test and therefore independent from other parts of the system as reported by Setiani \textit{et al.} \cite{Setiani2021251}. Test coupling and cohesion discussed by Palomba \textit{et al.} \cite{Palomba2016130} describe dependencies between tests and are included in this factor.

    \item \textbf{Comments (2)}: Single comments in test code providing useful information. According to Fisher and Johnson \cite{Fisher2018800}, one of the differences between their generated tests and human tests is the lack of explanatory comments. In Setiani \textit{et al.} \cite{Setiani2021251}, survey participants also mention comments being to some degree important to readability.

    \item \textbf{Textual features (1)}: Textual features focus on natural language properties part of test cases like consistency of identifiers or identifiers present in a dictionary. These features can be easily computed and are therefore frequently used in readability models and automatic readability raters like in Scalabrino \textit{et al.} \cite{scalabrino2016improving}.

\end{description}

\begin{framed}
    \noindent\textit{RQ1.1 Findings. Which influence factors are analyzed in scientific literature?} A total of 9 influence factors have been found in the scientific literature, which can be grouped into individual factors and readability models. The three most frequently  mentioned individual factors are \textit{test names}, \textit{assertions}, and \textit{identifier names}. Readability models combine several individual factors related to \textit{test structure} or \textit{textual features}.
\end{framed}

\subsubsection{Which research methods are used in scientific studies (RQ1.2)?}
\label{subsub:RQ1.2-results}
In  RQ1.2, we give an overview on the study types and the used methods.
This analysis is based on the classification of established empirical research methods involving human participants \cite{wohlin2012experimentation}.
Although software tools for investigating the readability of software code exist, the readability of software tests is not in the main focus of these approaches. 

Concerning the utilized types shown in \autoref{tab:selected-studies}, most studies (15) report an experiment which is combined with a survey in 7 studies. Human involvement is quite common, in 16 from 19 studies humans take part in experiments, surveys or play another role as participants of the study. Next, we present details on the individual types of studies.

\begin{description}
\item[\textbf{Experiment (15)}:] 12 of 15 studies evaluate the effect of an approach with humans by either asking participants to answer questions to a given test case or code snippet without knowing the origin like in Roy \textit{et al.} \cite{Roy2020287} or Daka \textit{et al.}~\cite{Daka201757} or participants have to choose between two versions (forced choice) like in Setiani \textit{et al.} \cite{Setiani2021251} or Daka \textit{et al.} \cite{Daka2015107}. Alsharif \textit{et al.}~ \cite{Alsharif2019SQL} enhance their experiment by letting some participants vocalize their thoughts while filling out a questionnaire in a Think Aloud Study. Li \textit{et al.} \cite{Li2018Aiding} do not fit in this categorisation. They use an indirect approach to measure the effect of generated tags by letting one group write summaries of test cases with and without treatment. Another group then rates these summaries according to a scheme. The difference in the ratings shows the effect of the treatment. For analysing the experiments results, eight from 15 studies use a form of the Wilcoxon test, most commonly the Wilcoxon rank sum test. Furthermore, these studies report the effect size with the Vargha-Delaney ($\hat{A}_{12}$) statistic or Cliff's Delta. Three of these studies also use the Shapiro-Wilk test for normal distribution to decide if a parameterized test can be applied. Alsharif \textit{et al.}~ \cite{Alsharif2019SQL} use a Fisher's Exact test on their results. The remaining studies interpret the results without statistical tests.

\item[\textbf{Survey (8)}:]
Five out of eight studies use online questionnaires, one uses an off-site questionnaire and for one study the kind of survey could not be extracted. In the surveys six out of eight studies use Likert scales often for rating readability. Free text answers are also common for optionally elaborating on a rating or as a mitigation against random readability ratings like in Daka \textit{et al.} \cite{Daka201757}.

\item[\textbf{User study} and \textbf{Prototype} (3):] The three studies of these types use surveys with Likert scale in Lin \textit{et al.} \cite{Lin2019204}, forced choice questions with opportunity to elaborate on the rating in Zhang \textit{et al.} \cite{Zhang2016625} or a mixture of multiple- and binary-choice and open questions in Li \textit{et al.} \cite{Li2018Aiding}. Zhang \textit{et al.} \cite{Zhang2016625} use the Wilcoxon test for comparing the results of a prototype tool with other tools after a test on normality with the Shapiro-Wilk test.

\item[\textbf{Concept Paper (1)}:] Bowes et al. \cite{Bowes20179} brainstorm and discuss quality evaluation of software tests with industry partners. Afterwards they merge the result with their own teaching experience and relevant scientific literature and books on software testing.

\end{description}

\begin{framed}
    \noindent\textit{RQ1.2 Findings. Which research methods are used in scientific studies?}
    For gathering humans opinion on readability online questionnaires with Likert scales and free text answers are common. The dominant result analysis consists of a statistical analysis with a Wilcoxon test after an optional test on normality with the Shapiro-Wilk test.
    
\end{framed}

\section{Grey Literature Review}
\label{sec:grey-literature}
In this section, we first describe the study protocol and process for the  grey literature analysis followed by presentation of the results including a discussion of the respective research questions. The data set is available online~\cite{dataset2023}.

\subsection{Study Protocol and Process}
\begin{figure}[ht]
    \centering
    \includegraphics[scale=0.95]{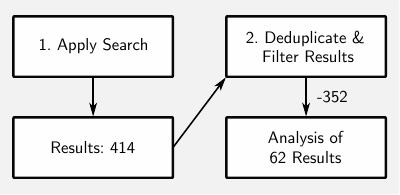}
    \caption{Grey literature review process and amount of received grey sources.}
    \label{fig:Grey-process}
\end{figure}

The process for conducting the review of grey literature (\autoref{fig:Grey-process}) is similar to the scientific literature review, except that there is no backward snowballing. The guidelines and recommendations by Garousi \textit{et al.} \cite{garousi2019guidelines} were used as input for this part of our work. We decided to add grey literature to this work, because testing is frequently performed by practitioners, and we assume that for them the internet is one of the first places used for information gathering and sharing. 

\paragraph{\textbf{Step 1: Apply Search.}}
Based on the research questions and knowledge obtained from the previous literature search we used the search strings \textit{``test code" readability} and \textit{``test code" understandability}. We performed these queries separately on \textit{Google} using a script for extracting all results. The script mimics a search without being logged in with a Google account. Therefore personalized search results should be reduced to some degree. In contrast to Googles prediction of hundreds of thousands of results, it returned 146 results for \textit{"test code" readability}, 101 for \textit{"test code" understandability} and 167 for \textit{"test code" legibility} (total: 414) in mid-February 2022.

\paragraph{\textbf{Step 2: Deduplicate \& Filter Results.}}
We first deduplicate the results by comparing the links which removed 18 sources. The result set was imported into a spreadsheet for applying inclusion and exclusion criteria.\\

\noindent \textbf{Inclusion Criteria.} We included a study if the following criterion was fulfilled:

\begin{itemize}
    \item Readability or understandability of test code is a relevant part of the source. This is the case if the length of the content on readability is sufficient and if the source contains concrete examples of factors influencing readability.
\end{itemize}

\noindent \textbf{Exclusion Criteria.}
We excluded a study if one of following criteria applied:
\begin{itemize}
    \item Not written in English
    \item Literature indexed by \textit{ACM}, \textit{Scopus}, \textit{IEEE}
    \item Duplicates, videos, dead links 
\end{itemize}

The criteria were evaluated based on the contents of the source. This step left us with 62 results ready for further analysis and extraction of influence factors. \\

\noindent\textit{Excluded Sources. }
Similar to the scientific literature search, we provide some examples and rationale for sources that where excluded when applying the defined criteria:
Source Karhik \footnote{Karhik, Use AssertJ to improve your test code readability ... - Upnxtblog, https://www.upnxtblog.com/index.php/2018/04/25/use-assertj-to-improve-your-test-code-readability-maintenance-of-tests-easier/} is a blog entry, which is relatively short and primarily lists features of AssertJ. Although the entry mentions readability improvement by using AssertJ in one sentence, it gives no explanation for this claim. Source Bas Broek\footnote{Bas Broek, (Improving Your) XCTAssert* Failure Messages - Bas' Blog, https://www.basbroek.nl/xctassert-asterisk} is a large blog entry with a primary focus on the readability of assertion failure messages. Factors relevant for our study (naming of test cases and test structure) are mentioned in only three sentences, which is a negligible small part of the whole entry. 
With the same rationale we also exclude source Wikipedia\footnote{Wikipedia, Test-driven development, https://en.wikipedia.org/wiki/Test-driven\_development}, because the primary focus is on test driven development and readability is a brief side topic. Likewise, Programmer All\footnote{Unit test 2 - Programmer All, https://www.programmerall.com/article/98141652585/} has much content and also provides code snippets. Still, the focus is on unit testing in general and not on improving readability. Source Karlo Smid\footnote{Karlo Smid, Kill The Unit Test - Tentamen Software Testing Blog, https://blog.tentamen.eu/kill-the-unit-test/} discusses the DRY-principle (don't repeat yourself) in context of unit testing. However, the blog entry is very short and primarily references to another source already present in the result set \cite{Derek_Snyder_and_Erik_Kuefler2019}. Although the collaborative source Openstack\footnote{TestGuide - OpenStack wiki, https://wiki.openstack.org/wiki/TestGuide} has a reasonable size and it also has a section on readability, the statements are too generic and do not contain a concrete influence factor on readability. 
Finally there are also many sources which are off-topic, e.g, because they discuss general code readability or quality, they describe advantages of unit testing, or they are documentation pages of test frameworks.

 \subsection{Grey Literature Analysis Results}
 \label{subsec:greyLiterature}
 In the following, we present the results from our further analysis of the grey literature sources with regard to factors influencing readability, and we provide answers to the research questions RQ2.1 and RQ2.2.
 
\textbf{Infuence factors.} We identified 12 types of influence factors in the analysis of the gray literature. The factors are related to \textit{test structures (Str)}, \textit{test names (TeN)}, \textit{assertions (Asse)}, \textit{helper structures (Help)}, \textit{dependencies (Dep)}, \textit{identifier names (IdN)}, \textit{fixtures (Fix)}, \textit{DRY principle (DRY)}, \textit{test data (TeD)}, \textit{comments (Com)}, \textit{domain specific language (DSL)}, and \textit{parameterized test (Par)}. A detailed description of each factor is provided in Section \ref{subsub:RQ2.1-results}.

The influence factors were extracted from the literature sources by one of the authors by tagging each source with keywords, which are mentioned in the context of test code readability. Keyword mentioned in a different context were not included. For example, in \cite{Jon_Reid2016} the use of ``helper methods" is only mentioned in context of easier maintenance, therefore this appearance of the factor helper methods is not counted. The results were cross-checked and discussed by the other authors of the study. 

In our analysis, we also investigated what types of gray literature sources we analyzed, when the literature sources mentioning the influence factors were published, and in context of what programming languages readability was discussed. 

\begin{figure}[tb]
    \begin{subfigure}[c]{0.49\textwidth}
        \centering
        \includegraphics[width=\textwidth]{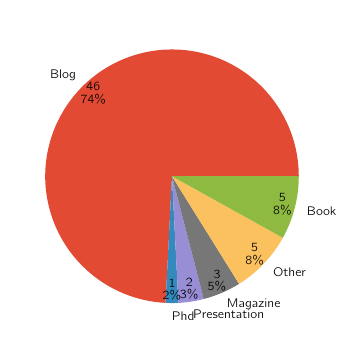}
     
    \end{subfigure}
    ~
    \begin{subfigure}[c]{0.46\textwidth}
    \centering
        \begin{tabular}{lll}
        \toprule
        \textbf{Type} & \textbf{Absolute} & \textbf{Relative} \\ \midrule
        Blog & 46 & 74\% \\
        Book & 5 & 8\% \\
        Other & 5 & 8\% \\
        Magazine & 3 & 5\% \\
        Presentation & 2 & 3\% \\
        Phd & 1 & 2\% \\
        \midrule
        \textbf{Sum:} & 62 & 100\% \\ \bottomrule
    \end{tabular}
    \end{subfigure}
    \caption{Types of analyzed selected grey sources.}
    \label{fig:grey-types}
\end{figure}

\textbf{Source types.}
\autoref{fig:grey-types} shows the identified types of grey literature source. From the 62 sources around 75\% (46 in total) are identified as blog entries of various sizes. A source is also identified as a blog when there is no clear indication that an editorial team is involved. The types of the remaining 16 sources are spread across 5 books, 5 other types (stackoverflow, quora, wiki, cheatsheet, podcast), 3 magazines, 2 presentations (slide shows), and 1 Phd thesis.

\begin{figure}[t]
    \centering
    \includegraphics[scale=1.0]{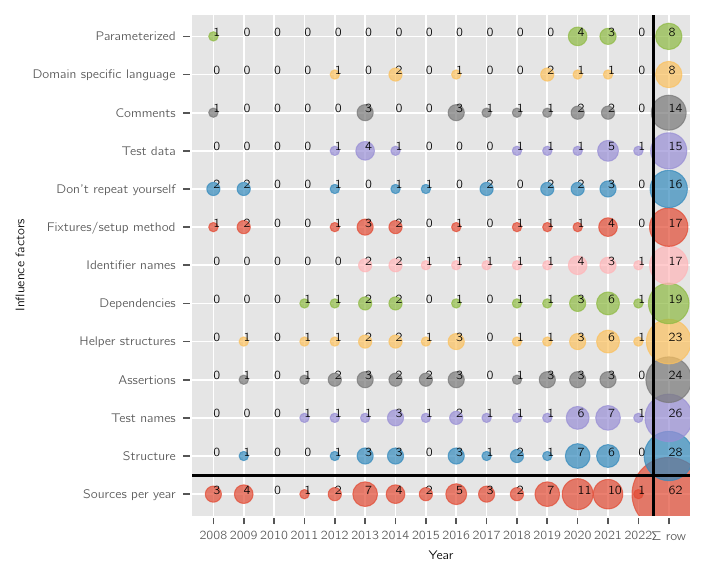}
    \caption{Factors investigated by grey literature. The bottom row gives the total number of sources per year, which may cover multiple factors.}
    \label{fig:grey-factors}
\end{figure}

\textbf{Factor across years.}
\autoref{fig:grey-factors} shows the factors investigated by the blogs across the years. The bottom line \textit{Sources per year} gives the number of sources in a particular year which investigated the factors above. Apart from parameterized tests, which appeared only seven times in the years 2020 and 2021 and in fewer sources in 2017 and 2018, there are no obvious fluctuations in the distribution of factors. \autoref{tab:grey-sources} shows the selected sources ordered by years descending and the investigated factors in detail, where these effects are also visible.

\textbf{Programming languages.}
Concerning programming languages, 19 sources mention Java or use Java code snippets, C\# appears in 10, Java Script in nine sources and, Ruby in three sources. Kotlin and Python each appear in two sources, Scala, Typescript, C++ and Go are mentioned in one source each. Some sources do not mention a certain programming language or do not use code snippets, because they provide general best practices for testing.
This is in accordance with the findings of our previous SMS~\cite{winkler2021vst}, where Java is the dominant language used in studies on test code readability.

\subsubsection{Which influence factors are discussed in grey literature (RQ2.1)?}
\label{subsub:RQ2.1-results}

In the following, each of the 12 influence factors identified in the gray literature analysis is described in detail. The number in brackets shows how many of the 62 reviewed literature sources mention the factor. They range from 28 (45\%) to 8 times (13\%). \autoref{tab:grey-sources} lists the analyzed grey literature sources (rows) and shows in which of these sources the identified influence factors (columns) are mentioned. For the sake of completeness, the table shows all 14 influence factors identified in the scientific as well as in the gray literature search, which includes two factors not mentioned in the gray literature.

\begin{table}
    \input{grey_literature_factors_caption}    	    
    \input{grey_literature_factors}

    \label{tab:grey-sources}
\end{table}

\begin{description}

\item \textbf{{Test structure} (28)} (Str):
23 out of 28 sources suggest the use of patterns like \textit{Arrange, Act, Assert} \cite{Arho_Huttunen2021}, \textit{Given, When, Then} \cite{Bas_Dijkstra2016} or \textit{Build, Operate, Check} \cite{Anmol_Sarna2018}. Two sources (\cite{Tuomas_Kareinen2012} and \cite{Tobias_Goeschel2016}) suggest to group similar test cases to see differences more quickly. Other sources \cite{Matheus_Rodrigues2018}\cite{Javier_Fernandes2020} suggest to watch out for ``eye-jumps", e.g., a variable, which is initialized many line breaks away from its usage. The absence of logic, shortness, and coherent formatting of test cases is also mentioned by several authors.

\item \textbf{{Test names} (26)} (TeN):
All sources suggest coherent naming of test cases and most of them suggest a concrete naming pattern like \textit{givenFooWhenBarThenBaz} \cite{Anshul_Bansal2021} or \textit{subject\_scenario\_outcome} \cite{Jenny_Shih2020}. Three sources (\cite{Arho_Huttunen2021}, \cite{Flavio_nickname_Company_blog2016} and \cite{Philip_Hauer2021}), explicitly suggest to use spaces in test names, which is a practice also shown by others in code examples, e.g., \cite{Jenny_Shih2020} and \cite{Peter_Bloomfield2020}. Long names are explicitly okay for two sources, since these methods are not called in other parts of the code. Different opinions exist on the inclusion of the name of the concrete tested method in the test name. 
\cite{Dustin_Boswell_Trevor_Foucher2012}, \cite{Petri_Kainulainen2014} and \cite{Matheus_Rodrigues2018} suggest to include the method name in the test name. Other sources like \cite{Pawel_Lipinski2013}, \cite{Henry_Coles_and_others2015} and \cite{Dan_Carter2011} do not recommend to include the method name, because if the method name changes the test name has to change too. Instead the tested behavior should be described.

\item\textbf{{Assertions} (24)} (Asse):
The use of appropriate assertions or custom assertions is suggested in eleven sources, e.g., \cite{Dan_Carter2011} and \cite{Anshul_Bansal2021}. Nine sources mention assertion libraries like AssertJ (Java) or FluentAssertions (C\#) since they enable a more natural language style for asserting properties and contain additional assertions for collection types \cite{Jason_Roberts2019}\cite{Bas_Dijkstra2016}. Four sources stress the importance of assertion messages for debugging. Concerning the amount of assertions, the rule ``one assertion per test" is mentioned by, e.g., \cite{Arho_Huttunen2021} and \cite{Tobias_Goeschel2016}.

\item \textbf{Helper structures (23)} (Help): 
13 sources recommend helper methods in order to hide (irrelevant) details like creating objects or asserting properties \cite{Hugh_Grigg2020}\cite{Marc_Duiker2016}. The \textit{Builder Pattern} (or similar patterns) are used by six sources for creating the objects under test, e.g., \cite{Mark_Needham2009}\cite{T._Yonekubo2021}. Inheritance of test classes is seen critically by some authors, e.g., \cite{Jan_Van_Ryswyck2021}\cite{Gil_Zilberfeld2014}.

\item \textbf{{Dependencies} (19)} (Dep):
All 19 sources agree that one test should only test one functionality or behavior. This affects readability positively, because the test stays short and the test name can be more descriptive, since only one behavior has to be described. Four sources highlight to only assert properties which are absolutely necessary for the functionality described by the test name and to resist the urge to check additional properties.

\item \textbf{{Identifier names} (17)} (IdN):
While nine sources only give generic information (e.g. \textit{should have meaningful or intention revealing names}), other sources provide detailed recommendations suggesting, e.g., to either prefix variables with \textit{expected} and \textit{actual} \cite{Philip_Hauer2021} or use names like \textit{testee, expected, actual} \cite{Henry_Coles_and_others2015}.

\item \textbf{Fixtures (17)} (Fix):
Although 15 of the 17 sources use fixtures, sometimes in combination with setup methods, two sources \cite{Gil_Zilberfeld2014}\cite{Matheus_Rodrigues2018} argue against the use of fixtures, because they are not visible in the test itself and may contain important information. Similarly, \cite{Philip_Hauer2021} points out that moving reusable test data into a fixture forces the reader to jump between two locations. Finally, \cite{Urs_Enzler2014} suggests that fixtures should only be used for infrastructure and not for the system under test, and \cite{Fernando_Chovich_Correa2019} recommends to use them only for properties which are needed in every test case.

\item \textbf{DRY principle (16)} (DRY):
In the sources which mention the \textit{\textbf{D}on't \textbf{R}epeat \textbf{Y}ourself} (DRY) principle, there is an agreement that strict adherence to this principle hides away information important for understanding test cases. Others favour the \textit{\textbf{D}escriptive \textbf{A}nd \textbf{M}eaningful \textbf{P}hrases} (DAMP) principle \cite{Henry_Coles_and_others2015} or to find a balance between these principles. As a combination of both, two sources \cite{Vladimir_Khorikov2008}\cite{Jan_Van_Ryswyck2021} suggest to clearly show what a test does (DAMP), but to hide how it is done in a helper method (DRY).

\item \textbf{{Test data} (15)} (TeD):
Five authors suggest to avoid literal test data (a.k.a. magic values), instead local variables, constants or helper functions should be used to provide additional information, e.g., \cite{Lasse_Koskela2013}\cite{Petri_Kainulainen2014}. However, \cite{Philip_Hauer2021} argues that declaring local variables for this purpose can quickly increase the test size and the mapping between variable and actual value has to be kept in mind when reading the test. Similarly, \cite{Fernando_Chovich_Correa2019} states that using literal values instead of variables sometimes improves readability.
Finally, test data should be production-like and simple, and one author also recommends to highlight important data.

\item \textbf{{Comments} (14)} (Com):
Eleven sources use comments in their snippets or mention them in the text to highlight \textit{Arrange, Act, Assert} or similar structures. However, this is not a strict rule for every author, e.g., source \cite{Jenny_Shih2020} uses empty lines as an alternative or \cite{Bas_Dijkstra2016} mentions to use comments with respect to the capabilities of the testing framework. If the framework already provides such structural hints then comments are unnecessary. Common code comments are mentioned by three sources with the general advice to avoid them, e.g., \cite{Tobias_Goeschel2016}.

\item \textbf{Domain specific language} (8) (DSL): 
In order to make tests more readable also for non programmers, some sources, e.g., \cite{Adit_Lal2019}\cite{NAIDELE_MANJUNATH_and_OLIVIER_DE_MEULDER2019}, suggest using helper functions or Gherkin (applied in Behavior Driven Testing with Cucumber) as domain specific languages. Such languages describe the executed behavior in natural language and, thus, hide the execution details.

\item \textbf{Parameterized test} (8) (Par):
Eight sources suggest to use parameterized tests (aka data-driven or table-driven tests) to reduce code duplication. This is also suggested by authors who are not in strict favor of the DRY principle, e.g., \cite{Philip_Hauer2021}.

\end{description}

\begin{framed}
\noindent \textit{RQ2.1 Findings. Which influence factors are discussed in grey literature?}
\label{par:RQ2.1-answer}
A total of 12 influence factors were found in the gray literature. They were mentioned from 28 to 8 times in the 62 analyzed sources. The five most often mentioned factors are related to test structures (28), test names (26), assertions (24), helper methods (23), and dependencies (19).
\end{framed}

\subsubsection{What is the difference between influence factors in scientific literature and grey literature (RQ2.2)?}
\label{subsub:RQ2.2-results}

From the total 12 factors identified in the grey literature review, 7 were already known from the scientific literature, while 5 factors were only found in the grey literature. These factors are new and did not appear in our previous white literature study~\cite{winkler2021vst}. In the scientific literature review we identified a total of 9 factors. It included two factors, which were mentioned only in the scientific literature but not in the gray literature. In the description below, the numbers in the brackets (\textit{A} vs \textit{B}) indicate how often a factor was found in the scientific literature versus in the grey literature.  

 However, even if factors have been found in both sources, the specific view on a factor can sometimes vary between white and gray literature. For example, quantifiable structural properties like line length or number of identifiers tend to be in the focus of scientific literature, whereas grey literature sources focus more on the semantic structure, e.g., the \textit{Arrange-Act-Assert} pattern. \autoref{tab:overlapping-factors} provides an overview of the differences identified in our analysis.

\begin{table}[t]
    \caption{Differences in influence factors between scientific and grey literature. (Overlapping aspects such as recommendations and discussions are highlighted.)}
    \input{overlapping_factors.tex}
    \label{tab:overlapping-factors}
\end{table}

\begin{description}

\item \textbf{{Test structure} (5 vs 28)} (Str):
Literature published in academic context tends to focus more on countable properties like maximum line length, amount of control structures, etc. (see, e.g., Grano \textit{et al.} \cite{Grano2018348}, Daka \textit{et al.} \cite{Daka2015107} or Setiani \textit{et al.} \cite{Setiani2020169036}). In contrast, the authors of grey literature sources focus on a semantic form of structure like the AAA pattern, which is also discussed in another study by Setiani \textit{et al.} \cite{Setiani2021251}. They report moderate positive influence on readability from this \textit{Arrange, Act, Assert} structure.

\item \textbf{{Test names} (6 vs 26)} (TeN):
Like in the grey literature, scientific literature  also mentiones the use of naming patterns, e.g., when test cases are renamed. Zhang \textit{et al.} \cite{Zhang2016625} or Daka \textit{et al.} \cite{Daka201757} use \textit{testSubjectOutcomeScenario} where ``Subject" is the method under test, although \textit{outcome} and \textit{scenario} can be left out. The approach by Roy \textit{et al.} \cite{Roy2020287} generates test names with a machine learning model based on the body of the test. According to examples given in the paper, this approach does not seem that it has to include the concrete method under test in the name. In other studies, e.g., by Panichella \textit{et al.} \cite{Panichella2016547} or Setiani \textit{et al.} \cite{Setiani2020169036}, survey participants highlight the importance of meaningful test names.

\item \textbf{{Assertions} (5 vs 24)} (Asse):
Some grey sources suggest to apply the ``one assertion per test" rule. However, there is little evidence in scientific literature about the effect of assertions on readability. Setiani \textit{et al.} \cite{Setiani2021251} report low influence from assertion messages on readability. Furthermore, Setiani \textit{et al.} \cite{Setiani2021251} and Daka \textit{et al.} report negligible influence from the amount of assertions. Studies like Bai \textit{et al.} \cite{BaiAssertionRoulette} or Panichella \textit{et al.} \cite{panichella2022test} from the field of test smells confirm the negligible importance of assertion messages and the number of assertions. Almasi \textit{et al.} \cite{Almasi2017263} report concerns from developers about the meaningfulness of generated assertions. Leotta \textit{et al.} \cite{Leotta2018AssertJ} report no significant influence on test comprehension when AssertJ is used instead of basic JUnit assertions. This contradicts the voices from grey literature, which suggest to improve readability with fluent assertions.

\item \textbf{Helper structures (0 vs 23)} (Help): 
This factor has been \textit{identified only in the grey literature}. In this context, the builder pattern and similar patterns are discussed, relating to practical recommendations for good design. 

\item \textbf{{Dependencies} (3 vs 19)} (Dep):
The recommendation that one test should only test one functionality or behavior is mentioned by Palomba \textit{et al.} \cite{Palomba2016130}\footnote{The recommendation itself was proposed by Van Deursen et al. \cite{van2001refactoring}.}. The participants from the study of Setiani \textit{et al.} \cite{Setiani2021251} to some extent agree that a unit test should only depend on one unit, which reflects the opinion of this factor from grey literature.

\item \textbf{{Identifier names} (5 vs 17)} (IdN):
The survey from Lin \textit{et al.} \cite{Lin2019204} shows the importance of meaningful, concise and consistent identifiers. The renaming approach by Roy \textit{et al.} \cite{Roy2020287} also suggests variable names like \textit{expected} and \textit{result}. Their deep learning model was trained with software projects of a high level of quality. Therefore, it seems plausible that identifier names as those mentioned in the grey literature sources are commonly used in tests of high quality projects. 

\item \textbf{Fixtures (0 vs 17)} (Fix):
This factor has been \textit{identified only in the grey literature}, which discusses arguments for and against the use of test fixtures from a practical perspective. 

\item \textbf{DRY principle (0 vs 16)} (DRY):
This factor has been \textit{identified only in the grey literature}, n context of practical recommendations on how to apply this development principle to test code.

\item \textbf{{Test data} (4 vs 15)} (TeD):
Participants of the workshop from Bowes \textit{et al.} \cite{Bowes20179} also recommend to avoid magic values. Almasi \textit{et al.} \cite{Almasi2017263} and Afshan \textit{et al.} \cite{Afshan2013352} highlight importance of meaningful or human-like test data.

\item \textbf{{Comments} (2 vs 14)} (Com):
The usage of comments for highlighting the structure of the test is not investigated in scientific literature. Fisher and Johnson \cite{Fisher2018800} explain different readability ratings between generated tests and human tests also with the lack of explanatory comments. Setiani \textit{et al.} \cite{Setiani2021251} survey participants who also mention comments being to some degree important to readability. These findings are to some extent contradicting the recommendation in grey literature, which is generally to avoid such explanatory comments.

\item \textbf{Domain specific language} (0 vs 8) (DSL): 
This factor has been \textit{identified only in the grey literature}. It relates to test frameworks used in practice such as Gherkin.

\item \textbf{Parameterized test} (0 vs 8) (Par):
This factor has been \textit{identified only in the grey literature}. It relates to practical suggestions to reduce code duplication by using parameterized tests.

\item \textbf{Test summaries (4 vs 0)} (TS):
This factor has been \textit{identified only in the scientific literature}. It is related to the application of source code summarization techniques investigated in related research as support for understanding test code. 

\item \textbf{Textual features (1 vs 0)} (TF):  
This factor has been \textit{identified only in the scientific literature}. It is related to the application of natural language processing investigated in related research for test cases.
\end{description}

\begin{framed}
\noindent \textit{RQ2.2 Findings. What is the difference between influence factors in scientific literature and grey literature? }
\label{par:RQ2.2-answer}
9 factors were identified in the scientific and 12 factors in the gray literature review, with an intersection of 7 factors identified in both. The factors identified most often in the scientific literature were also most frequently found in the grey literature: Test structures (5 and 28 times), test names (6 and 26 times), and assertions (5 and 24 times). Scientific and grey literature sources sometimes focus on different aspects of common factors.  
\end{framed}

\section{Evaluation of Influence Factors}
\label{sec:experiment}

For the following experiment we take the results from the systematic mapping study (Section~\ref{sec:sms}) and the grey literature review (Section~\ref{sec:grey-literature}) and investigate a selection of identified influence factors with focus on the perception of test case readability.
\begin{figure}[ht]
    \centering
    \includegraphics[scale=0.85]{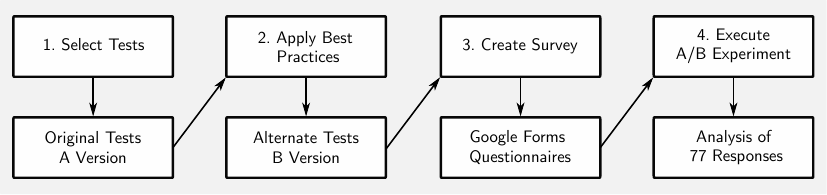}
    \caption{Experiment process and amount of received responses.}
    \label{fig:challenge-process}
\end{figure}

\subsection{Experiment Setup and Procedure}

The experiment follows an A/B testing approach. The participants rate readability of original and altered test cases. Experiments based on A/B testing are a good approach for comparing the effect of a treatment to a population. In our scientific literature review we also found some studies using this approach, e.g., Roy \textit{et al.} \cite{Roy2020287NormalBib} or Setiani \textit{et al.} \cite{Setiani2021251NormalBib}. 
Participants of a master course on software testing at TU Wien were invited to participate voluntarily in this online experiment with the possibility of bonus points for the course as a reward.
\autoref{fig:challenge-process} shows an overview on the experiment process. We discuss the individual steps in the following sections.

\begin{table}[ht]
    \caption{Listing of test cases with their assigned influence factor, originating project and differences made for both versions. \textbf{A} (original version) and \textbf{B} (altered version) denote the groups.}
    \input{test_cases_origin_factors.tex}
    \label{tab:test-cases-origin-factors}
\end{table}

\subsubsection{Select Tests}
\label{lab:exp-question}
We searched open source repositories for test case that are related to the influence factors we identified in our literature study, specifically test cases adhering or contradicting to these factors. We selected 30 test cases covering different influence factors from 8 sources, which also contain generated tests by Randoop and Evosuite. \autoref{tab:test-cases-origin-factors} shows influence factor, test name and origin project. Most tests including the automatically generated tests come from the open source project \textit{Apache Commons Lang3}. Other sources include the \textit{Spring Framework}, \textit{IntelliJ}, and \textit{Apache Flink}. The last three tests with origin project ``Student Solution" are selected tests written by students for a course assignment.

The test cases we found in our search and which are used in the subsequent experiment cover 7  out of the 14 influence factors (see \autoref{tab:grey-sources} in  \autoref{subsec:greyLiterature} for a complete list of influence factors),  identified in the literature study, since we limited our selection to only those test cases retrieved from real-world open source projects that can be clearly related to individual influence factors. Therefore, we included test cases related to the influence factors \textit{Structure (Str), Assertions (Asse), Dependencies (Depe), Test Data (TD), Comments (Co), Fixtures (Fix),} and \textit{Parametrized Tests (Para)} and excluded test cases related to \textit{Test Names (TeN), Identifyer Name (IdN), Test Summaries (TS), Textual Features (TF), Helper Structures (Help), DRY Principle (DRY),} and \textit{Domain Specific Language (DSL).}

\subsubsection{Apply Best Practices}
\label{lab:main-challenge}
For each test, we create an alternative version, following the findings from the literature study. For example, long test cases (variant A) were modified by splitting them up into two or more smaller test cases (variant B). This modification corresponds to the best practice suggested in \cite{Philip_Hauer2021}. Similarly, test cases using standard assertions (variant A) were modified by using dedicated assertion frameworks such as AssertJ (variant B), as suggested in \cite{Bas_Dijkstra2016}.

\autoref{tab:test-cases-origin-factors} provides an overview of the covered influence factors from the literature study and a short description on the differences between A and B version of the different tests in column \textit{``Modification A/B"}.

Furthermore, we used three additional test cases (not shown in the table) without modification as a control group. The purpose of these control tests is to verify that the participants show a consistent rating behavior for A and B tests, which allows us to assert the internal validity of the experiment.

\subsubsection{Create Survey}

We created surveys containing a subset of 12 test cases out of the entire set of the 30 tests listed in Table 6. Each survey contained an equal mix of A and B variants.
In total we created 6 different surveys to provide full coverage of all 30 tests in each of the variants.
The surveys were randomly distributed to the participants taking part in the experiment, who were unaware of the covered influence factors and whether the included tests were modified or unmodified.

The participants were asked to rate the readability on a 5 point Likert scale from 1 (unreadable) to 5 (easy to read) and to provide up to three free text reasons for their rating. Before and after this main task of the experiment, there is a pre- and post-questionnaire for collection information on the participants' background and feedback about the experiment run.

We developed the questionnaires using \textit{google.forms}, which provides an easy way for creating surveys that can also be reused for future replications. The collected data can be exported in various formats for processing and analysis. Beside the survey forms, we provided the selected tests in a PDF and as plain text files as additional supporting materials for the study participants.

\subsubsection{Execute A/B Experiment}
 The online survey was open for two weeks and the participants were free to start and stop their run at any time in this period. The duration for taking part in the experiment was about one hour per participant. In total, 77 participants completed the survey.
 
\subsubsection{Analysis}
 We use the software R to calculate the significance of the results with statistical tests on level of $\alpha=0.05\%$. According to an analysis with the Shapiro-Wilk test, the rating data does not follow a normal distribution. Therefore, and since our data is unpaired, we use the Wilcoxon Rank Sum test. When a significant difference between the distribution of the groups \textbf{A} and \textbf{B} is found, we report the effect size with Cliff's Delta ($\delta$). Roy \textit{et al.} \cite{Roy2020287NormalBib} used the same approach for their Likert scale data. Cliff's Delta is interpreted according to Romano \textit{et al.} \cite{romano2006appropriate} with $|\delta|<0.147$ \textit{``negligible",} $|\delta|<0.33$ \textit{``small"}, $|\delta|<0.474$ \textit{``medium"}, otherwise \textit{``large"}
 
 \subsection{Experiment Results}
 This section presents the results of the experiment on the readability of the selected set of test cases to investigate the related influence factors. 
 Some factors influencing readability appear more than once in \autoref{tab:test-cases-origin-factors} and the modifications have different goals. Therefore we analyse the differences between groups A and B across these modifications. 
We discuss each modification after an overview on the participants in the following sections.

\begin{table}[ht]
    \caption{Information on participants experience.}
    \centering
    \begin{subfigure}[c]{0.48\textwidth}
    \centering
        \caption{General Software Development Experience {[years]}.}
        \input{general_software_development_exp_years}
    \end{subfigure}
    ~
    \begin{subfigure}[c]{0.48\textwidth}
    \centering
        \caption{Professional Software Development Experience {[years]}.}
        \input{professional_software_development_exp_years.tex}
    \end{subfigure}
    \label{tab:participants-experience}
\end{table}

\textbf{Participants experience. }
To gather some information about our participants we asked for their amount of experience in general and professional software development in years. They could choose between 0, 1-2, 2-5 and \textgreater{}5 years. \autoref{tab:participants-experience} shows results of both questions. Almost 45\% of our participants have more than five years of experience in software development and more than 50\% have two to five years of experience. Concerning professional development around 30\% have either one to two or two to five years of experience. In total around 75\% have worked at least one year.

\subsubsection{Do factors discussed in practice show an influence on readability when scientific methods are used (RQ3.1)?}
\label{subsub:RQ3.1-results}
\autoref{fig:aggregated-ratings} shows the distribution of the aggregated readability ratings including boxplots for the investigated modification mapping to influence factors. \autoref{tab:stat-analysis} shows the results from the statistical analysis. The first column ``Modification A/B (Influence Factor) maps to the according columns in \autoref{tab:test-cases-origin-factors}. We discuss each modification in the following sections. As a reminder, we interpret Cliff's Delta ($\delta$) according to Romano \textit{et al.} \cite{romano2006appropriate} with $|\delta|<0.147$ \textit{``negligible",} $|\delta|<0.33$ \textit{``small"}, $|\delta|<0.474$ \textit{``medium"}, otherwise \textit{``large"},

\begin{description}
\item{\textbf{Loops vs. Unrolled (\autoref{fig:loops-unrolled}}).}
\label{para:loops-vs-unrolled}
In this modification the difference between A and B of the aggregated results is significant with $p=0.02$. The effect size $\delta=-0.35$ is on the lower end of a \textit{``medium"} effect size. The analysis of the individual tests reveals that the whole modification is significant, because of the last test with $p=0.01$ and $\delta=-0.67$ (\textit{``large"} effect). In this test the code contains two 2D arrays, nested loops to perform the test and string concatenation for the assertion message. The modified version primarily consists of assertions for all cases the loops generate, without assertion messages.

\item{\textbf{Try Catch vs. AssertThrows (\autoref{fig:trycatch-assertthrow}}).}
\label{para:try-catch-vs.-assertThrows}
Overall there is no significant difference in the readability ratings between the original and the modified versions. Only in the second test the difference between A and B is barely significant $p=0.04$, although it has a \textit{``large"} effect size with $\delta=-0.54$. One possible explanation for this result could be the relative short size of this test in comparison to the other ones in this modification. Due to the short length, there may be no possibility for other bad practices to mask the positive influence of this modification.

\item{\textbf{Variable Re-Use (\autoref{fig:variable-re-use}}).}
Neither the figure nor the statistical analysis show a significant difference in the ratings.

\item{\textbf{Structure (\autoref{fig:structure}}).}
Overall there is a clear difference between the groups of this modification with $p=0.0$ and a \textit{``large"} effect, $\delta=-0.59$. Only for one of the three tests the difference between groups is not significant with $p=0.16$.

\item{\textbf{Comments (\autoref{fig:comments}}).}
Although none of the individual tests has a significant difference between A and B, the aggregated result is significantly different with $p=0.02$ and has a lower \textit{``medium"} effect size with $\delta=0.36$. Since we removed comments in the original versions of the tests, the A version contains more information than B. A look at \autoref{fig:comments} and the median values in the \autoref{tab:stat-analysis} shows that the participants gave the A version better ratings. This is also reflected by the positive sign of the effect size. The comments do not highlight the structure of the test, they are of the nature ``explanatory comments". This is a confirmation of the positive influence of comments on readability found by scientific literature.

\begin{figure}[H]
    \centering
    \begin{subfigure}[c]{0.32\textwidth}
        \includegraphics[width=1\textwidth]{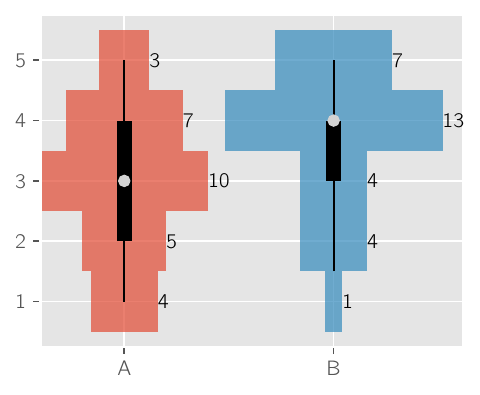}
        \caption{Loops vs. Unrolled}
        \label{fig:loops-unrolled}
    \end{subfigure}
        \begin{subfigure}[c]{0.32\textwidth}
        \includegraphics[width=1\textwidth]{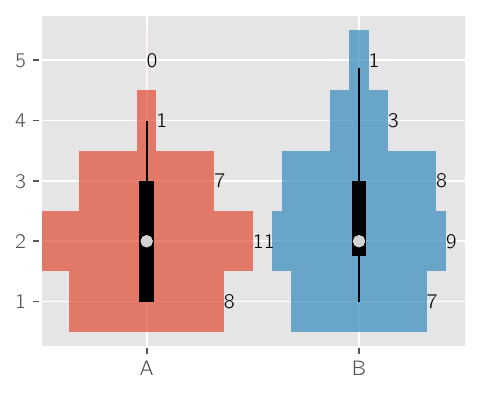}
        \caption{Try Catch vs. Asserts}
        \label{fig:trycatch-assertthrow}
    \end{subfigure}
        \begin{subfigure}[c]{0.32\textwidth}
        \includegraphics[width=1\textwidth]{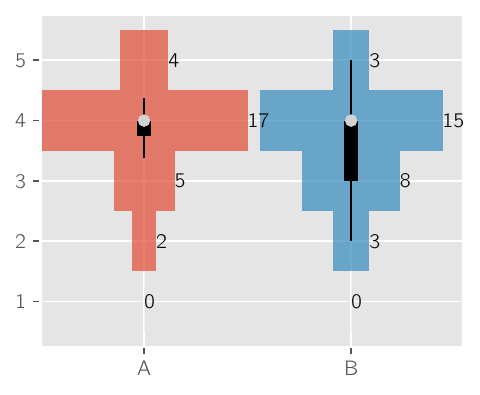}
        \caption{Variable Reuse}
        \label{fig:variable-re-use}
    \end{subfigure}
            \vspace*{\floatsep}
            \begin{subfigure}[c]{0.32\textwidth}
        \includegraphics[width=1\textwidth]{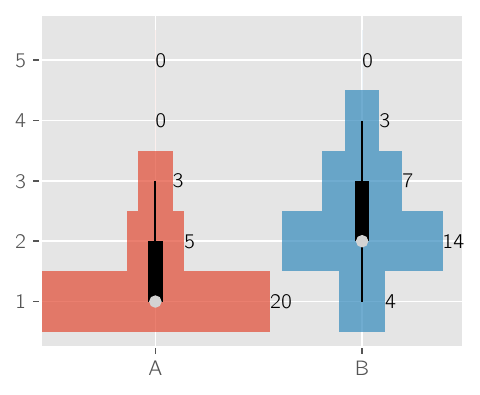}
        \caption{Package Names, If-Str.}
        \label{fig:structure}
    \end{subfigure}
        \begin{subfigure}[c]{0.32\textwidth}
        \includegraphics[width=1\textwidth]{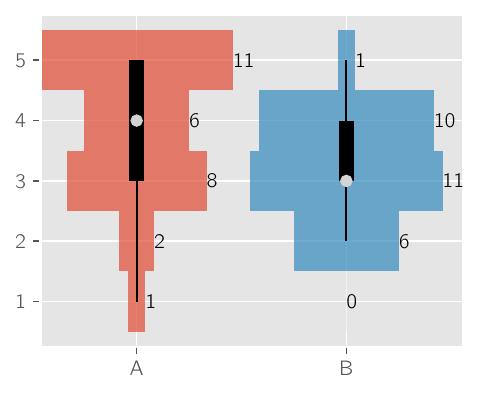}
        \caption{Remove Comments}
        \label{fig:comments}
    \end{subfigure}
        \begin{subfigure}[c]{0.32\textwidth}
        \includegraphics[width=1\textwidth]{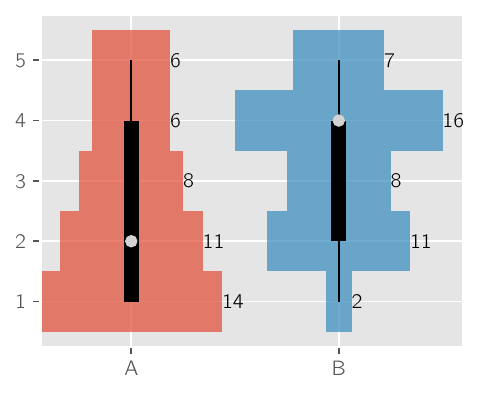}
        \caption{Loops vs. Parameterized}
        \label{fig:loops-paramet}
    \end{subfigure}
            \vspace*{\floatsep}
            \begin{subfigure}[c]{0.32\textwidth}
        \includegraphics[width=1\textwidth]{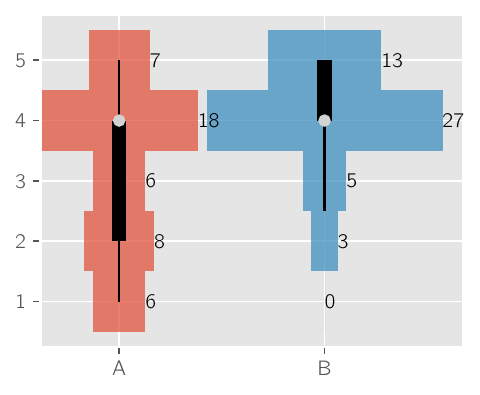}
        \caption{Split Up Tests}
        \label{fig:split-up}
    \end{subfigure}
        \begin{subfigure}[c]{0.32\textwidth}
        \includegraphics[width=1\textwidth]{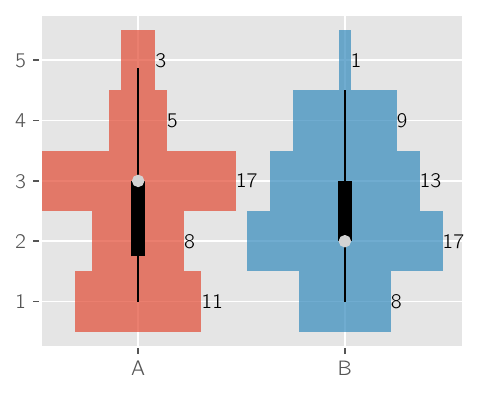}
        \caption{Specific Assertion}
        \label{fig:assertion}
    \end{subfigure}
        \begin{subfigure}[c]{0.32\textwidth}
        \includegraphics[width=1\textwidth]{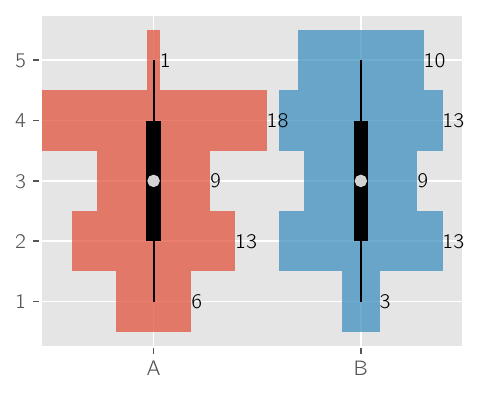}
        \caption{Unnecessary Try Catch}
        \label{fig:unnecessary-trycatch}
    \end{subfigure}
            \vspace*{\floatsep}
            \begin{subfigure}[c]{0.32\textwidth}
        \includegraphics[width=1\textwidth]{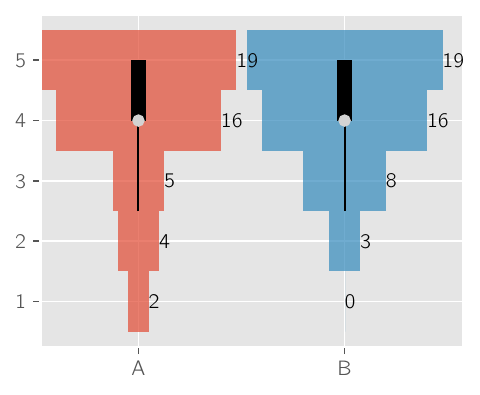}
        \caption{Remove Fixture}
        \label{fig:fixture}
    \end{subfigure}
    \caption{Distribution and box plots of aggregated readability ratings per A/B modification. Ratings from a five-point Likert scale range from 1 (not readable) to 5 (very readable). The numbers on the right hand side of the histograms represent the amount of answers for this rating.}
    \label{fig:aggregated-ratings}
\end{figure}

\begin{table}[H]
    \centering
    \caption{Statistical analysis of experiment results using a two-sided Wilcoxon Rank Sum test ($p$) and Cliff's D ($\delta$) for effect size. $\delta$ is only shown for $p<0.05$.}
    \input{stat_analysis_likert.tex}
    \label{tab:stat-analysis}
\end{table}

\item{\textbf{Loops vs Parameterized (\autoref{fig:loops-paramet}}).}
Like in \textit{Loops vs. Unrolled} the difference of the complete modification between groups A and B is significant with $p=0.00$ and $\delta=-0.34$, a lower \textit{``medium"} effect size, because of the last test. The original version is the same as in \textit{Loops vs. Unrolled} but the modified version extracts the test case data into an inlined CSV as input for the parameterized test case. The other forms of parameterized tests did not lead to significant changes in the readability ratings. 
In pursuit of the hypothesis from \autoref{subsub:control-groups} we also compare the ratings of this A group with the A group from \textit{Loops vs. Unrolled}. When looking at the median values the hypothesis seems to hold, because the values from this modification are lower in two of three tests. However, the Wilcoxon test does not detect a significant difference in the ratings with $p=0.11$.

\item{\textbf{Split Up (\autoref{fig:split-up}}).}
There is a clearly significant difference between A and B with $p=0.0$ but only a \textit{``small"} effect size, although with $\delta=-0.33$ it is on the edge to a ``medium" effect size. In detail there is one significant test $p=0.01$ with $\delta=-0.50$, a large effect size. When looking at the median values and the figures, we see that both versions are quite readable but the modified tests have few to no ratings in the lower part of the readability scale. 

\item{\textbf{Assertions (JUnit, Hamcrest, AssertJ) (\autoref{fig:assertion}}).}
There is no significant difference in readability when using standard JUnit assertions compared to assertions with Hamcrest or AssertJ assertions. This result confirms findings from Leotta \textit{et al.} \cite{Leotta2018}. 

\item{\textbf{Unnecessary Try Catch (\autoref{fig:unnecessary-trycatch}}).}
One test shows a significant difference with $p=0.01$ and $\delta=-0.53$, a \textit{``large"} effect size. With medians of $0.75$ the first test is almost very readable in both versions. However, we accidentally introduced an error in the modified version (we declared a variable twice, which is not allowed in Java). In the comments the participants noticed this error, therefore this error might mask the positive effect of the intended modification. The second test with medians of $0.25$ has a very long test name which the participants criticise. This again might mask the positive effect of the modification.

\item{\textbf{Fixture (\autoref{fig:fixture}}).}
We do not see a significant difference between the two versions neither in the figure nor in the table. The tests all have a quite good rating, which is could be caused by the participants knowledge about the system under test.
\end{description}

\begin{framed}
    \noindent\textit{RQ3.1 Findings. Do factors discussed in practice show an influence on readability when scientific methods are used?}
    \label{para:RQ3.1_answer}
    Applying industry best practices is no silver bullet for improving the readability of test code. 
    In the scientific experiments, the modifications showed a statistically significant positive influence on the readability of the tests for 50\% of the factors, i.e., in five out of ten cases. 
\end{framed}
\section{Summary, Threats to Validity, and Future Work}
\label{discussion}
This section summarizes the findings, discusses limitations and threats to validity, and provides future research directions.

\subsection{Summary}
The main goal of this paper was to combine scientific and practical views on the readability of software test code. We have conducted a Systematic Mapping Study (SMS) to cover relevant publications from academia to capture the scientific view on the readability of software test code. We have complemented the results of the SMS by taking into consideration practical views based on grey literature. Based on identified influence factors on test code, we conducted a controlled experiment in academic setting to explore the  perception of software test code readability with a set of 77 participants.

    We have identified \textit{unique readability factors in scientific literature} that include  
    readability models, application of code summaries used on test code that have been proposed and evaluated (see Section~\ref{subsub:RQ1.1-results}).
    
    \textit{Individual Influence Factors} have been evaluated in scientific literature by using scientific methods, such as online questionnaires with Likert scales and statistical analysis (see Section~\ref{subsub:RQ1.2-results}).

    \textit{Differences in scientific and grey literature.}
    In contrast to scientific literature grey literature provides a wide spectrum of best practices and guidelines concerning the influence factors of readability of test code. There is large overlap between science and grey literature, however in the overlapping factors we observed different views in the interpretations. (see Section \ref{subsub:RQ2.2-results}).
    
     \textit{Unique readability factors in grey literature.}
          Furthermore we found five additional factors exclusive to grey literature, e.g. helper structures, test fixtures. Concerning these factors there exist different views and even conflicting opinions, often related to the used/applied technology, testing framework, and test level/approach (see Section \ref{subsub:RQ2.1-results}).
    
    \textit{Empirical study of influence factors widely discussed in practice.}
    For half of the investigated modifications (Loops vs. Unrolled Loops; Package Names, If-Structures; Remove Comments; Loops vs. Parameterized and; Split Up Tests), which map to readability factors, we could show a statistical significant influence in test code readability. (see Section \ref{subsub:RQ3.1-results}). Other factors are less clear, which can be attributed to the nature of best practices, which are sometimes only applicable in specific contexts and not in general (e.g., modification Try Catch vs. AssertThrows; (see Section \ref{para:try-catch-vs.-assertThrows}).

\subsection{Limitations and Threats to Validity}
In this section, we summarize important limitations and threats to validity in context of the literature review (i.e., scientific and grey literature) and the empirical study.

\paragraph{Internal Validity}
\begin{itemize}
    \item In context of the Systematic Mapping study, the keyword, search string, analysis items, and the data extraction and analysis has been executed by one of the authors and intensively reviewed and discussed within the author team and external experts.
    \item The controlled experiment setup has been initially executed in a pilot run to ensure consistency of the experiment material. We have used a cross-over design of test case samples to avoid any bias of the experiment participants.
    \item  \label{subsub:control-groups} Three unmodified test cases were used as control groups in A/B testing. The Wilcoxon Rank Sum test does not suggest a significant difference between ratings provided by participants, when comparing groups with the same questionnaire. However, there is a significant effect when comparing control groups of different questionnaires. These results confirm a consistent rating behavior within groups and the significant differences between groups is as expected due to the independent ratings of participants from different groups.
\end{itemize}

\paragraph{External Validity}
\begin{itemize}
    \item We have conducted a literature reviews based on the guidelines of Petersen~\textit{et al.}~\cite{petersen2015guidelines} complemented by a systematic analysis of grey literature~\cite{garousi2019guidelines}. Therefore, the analysis results identified most prominent research directions in scientific literature complemented by practical discussions in non-academic sources (such as blogs). This approach enabled us to identify similar and/or different key topics in academia and industry.
    \item Experiment participants were recruited on a voluntary basis from three classes of a master course on \textit{software testing} at TU Wien. We captured background knowledge of the participants to identify participant experience. Most of the participants work in industry and can be considered as ``junior professionals". Therefore, the results are applicable for industry applications. 
    \item We used real-world test cases from open source projects as well as results from software testing exercises to ensure close to industry test cases. 
\end{itemize}

\paragraph{Construct Validity}
\begin{itemize}
    \item We build on best-practices for the literature review for academic publications  ~\cite{petersen2015guidelines} and grey literature~\cite{garousi2019guidelines} and followed experimentation guidelines, proposed by Wohlin \textit{et al.}~\cite{wohlin2012experimentation} for conducting the empirical study.
    \item For the controlled experiment, we captured individual test case assessments for A-B tests (i.e., original tests taken from existing projects and slightly modified test cases) based on a 5-point Likert scale.
    \item To avoid a bias introduced by the order of questions for the experiment, we reversed the question ordering for half of the experiment groups.
    \item To avoid random readability ratings, we asked participants to give reasons for their ratings as free text. Furthermore, the participants were told that their reward (bonus points) is coupled with active participation in the challenge.
    \item We tried to select test cases for A/B testing in our experiment, which could be clearly related to individual influence factors. Since the test cases we used were retrieved from real world projects instead of constructed examples, which could limit the relevance of our results, we only covered 7 out of the 14 influence factors identified in the literature search. Nevertheless, a certain amount of fuzziness with respect to influence factors may still be present, e.g., as discussed in the results for the modification Try Catch vs. AssertThrows (see \autoref{para:try-catch-vs.-assertThrows}).
    
\end{itemize}

\paragraph{Conclusion Validity}
\begin{itemize}
    \item We used the Shapiro-Wilk test for testing for normality, which would allow us to use a parametric statistical test. This approach is also used by Roy \textit{et al.} \cite{Roy2020287NormalBib} whose methodology is similar to ours.
    \item We used the non parametric Wilcoxon Rank Sum test, because our groups are unpaired and the Shapiro-Wilk test does not suggest a normal distribution of our result data.
    \item We report the effect size with Cliff's Delta, because it allows an interpretation of the magnitude of difference between two groups. It is also used by other studies in this field like Grano \textit{et al.} \cite{Grano2018348NormalBib}. 
    
\end{itemize}

\subsection{Implications for Research and Practitioners}
This section summarizes the main implications of the SMS (Section~\ref{sec:sms}), the grey literature study (Section~\ref{sec:grey-literature}), and the experiment (Section~\ref{sec:experiment}). \\

\paragraph{Implications for Research}
\begin{itemize}
    \item For the \emph{Software Testing} community, we identified influencing factors, observed only in grey literature, that could initiate additional research initiatives with focus on topics that are of interest for practitioners with limited attention of researchers.
    \item Researchers in \emph{Software Engineering} and/or \emph{Software Testing} can take up the results from literature review with focus on replicating and extending the presented research work.
    \item The \emph{Empirical Research} community can build on the the SMS protocol, the grey literature protocol, and the study design to replicate and extend the study protocol in different context.
    \item We selected a representative set of test cases that could be used by researchers to (i) design and develop a method and or tool to semi-automatically assess the readability of test code and (ii) to apply the test code set for evaluation purposes in different contexts.
     \item In the \emph{Software Testing} communities, factors, such as \emph{Setup methods/Fixtures}, \emph{Helper Methods}, \emph{DRYness} are widely discussed in the domain of practitioners. Considering these in test code generation could be useful for generating more readable tests. In a recent study Pannichella \textit{et al. } \cite{panichella2022test} also suggest to include capabilities for complex object instantiation into test suite generators.
    \item Finally, the findings of the study can be used as input for researchers from \emph{Software Engineering} communities to improve software maintenance tasks that benefit from readability assessments.

\end{itemize}

\paragraph{Implications for Practitioners}
\begin{itemize}
    \item For \emph{Software and System Engineering} organizations, results of this work can support software testers and developers to improve test code readability based on guidelines and identified influencing factors.
    \item \emph{Project and Quality Managers} can use the results to setup organization specific development guidelines to support software development, software testing, and software maintenance and evolution by a team of software experts. Applied best practices might  help to improve the quality of test cases and reduce effort and cost for maintenance activities.
    \item Factors with similar views from practitioners and academia include \emph{Test Names}, \emph{Identifier Names}, and \emph{Test Data}. For test and identifier names both domains agree on the use of naming patterns in order to achieve consistency across the test suite. For test data also both domains agree on the use of realistic and simple values and avoiding magic values.
    \item However, the experiment results show that application of best-practices is no guarantee for improved readability.
\end{itemize}

\subsection{Future Work}
The main goal of this article was to \textbf{Investigate the readability of Software Test Code} by combining scientific and practical views. We applied a systematic mapping study for analyzing scientific literature complemented by grey literature. Furthermore, we executed a controlled experiment in a Software Testing Master Course on academic level to investigate practical implications of a selected but typical set of test cases.

In the future, we plan to replicate the experiment to increase the external validity of the study in academia, complemented by industry participants. Furthermore, we plan to develop and evaluate a maturity model for the readability of test cases that could help quality managers, software test engineers, and software developer in better assessing the quality of test cases (from readability perspective) to improve software maintenance and evolution.

\section*{Acknowledgements}
This work has been partially supported via SBA Research (SBA-K1), a COMET Center within the COMET - Competence Centers for Excellent Technologies Programme, funded by BMK, BMAW, and the federal state of Vienna and managed by the Austrian Research Promotion Agency (FFG).
This work has been partially supported via the COMET Center “Austrian Competence Center for Digital Production” (CDP) [881843] and the K2 centre InTribology [872176].
This work has been partially supported by the Austrian Research Promotion Agency (FFG) in the frame of the project ConTest [888127] and via the COMET competence center INTEGRATE [892418] of SCCH, funded by BMK, BMAW, and the federal state of Upper Austria.
Finally, the financial support by the Christian Doppler
Research Association, the Austrian Federal Ministry
for Digital \& Economic Affairs and the National
Foundation for Research, Technology and Development 
is gratefully acknowledged.

\section*{Data Availability Statement}
The datasets generated during and/or analysed during the current study are available via the TU Wien research data respository ~\url{https://doi.org/10.48436/w4q8v-28695}.

\section*{Declarations}
\textbf{Conflicts of interest/competing interests:} None

\printbibliography[keyword=primary, title={References}, resetnumbers=true]

\appendix
\section*{Appendix}

\newrefcontext[labelprefix=G]
\printbibliography[keyword=grey, title=A) Sources of Grey Literature for Systematic Mapping, resetnumbers=true]

\newrefcontext[labelprefix=A]
\printbibliography[keyword=academic, title=B) Sources of Academic Literature for Systematic Mapping, resetnumbers=true]

\newpage
\section*{Author Biographies}
\begin{figure}[H]
    \includegraphics[width=0.3\textwidth]{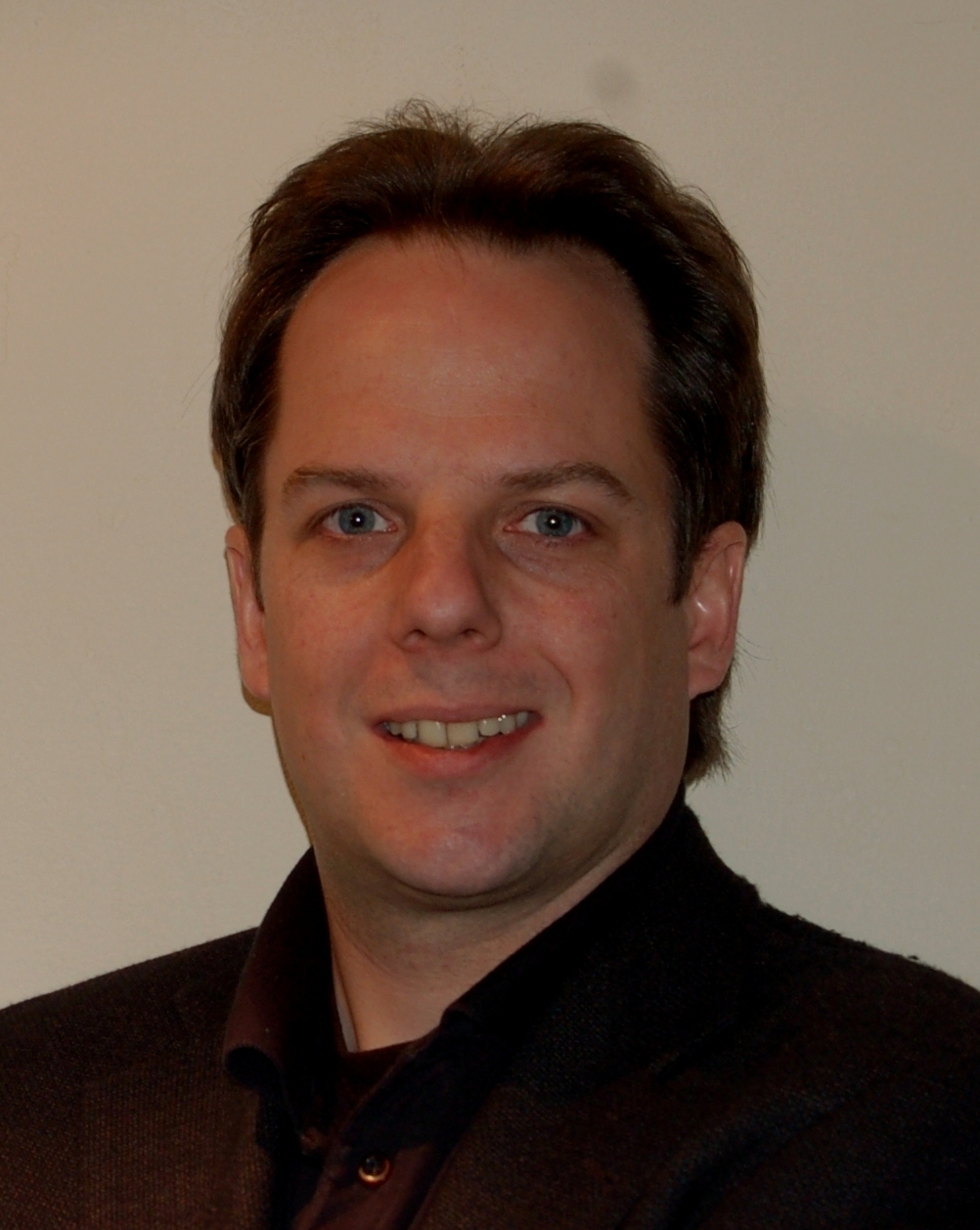}
\end{figure}
\textbf{Dietmar Winkler} is senor researcher at SBA Research, area manager at the Austrian Center for Digital Production (ACDP), and lecturer at the TU Wien, Vienna, Austria as member of the research group for Quality Software Engineering at the Institute for Information Systems Engineering in the Software Engineering Group. Dietmar holds an PhD and a M.Sc. in Computer Science from TU Wien, Austria. He has more than 20 years of experience in applied research in the fields of software and production systems engineering, engineering process improvement, quality assurance and testing, and empirical software engineering. He (co-)authored more than 130 peer-reviewed publications, (co-)organized international conferences, tracks, and sessions, and served as program committee member for international journals and conferences.

\begin{figure}[H]
    \includegraphics[width=0.3\textwidth]{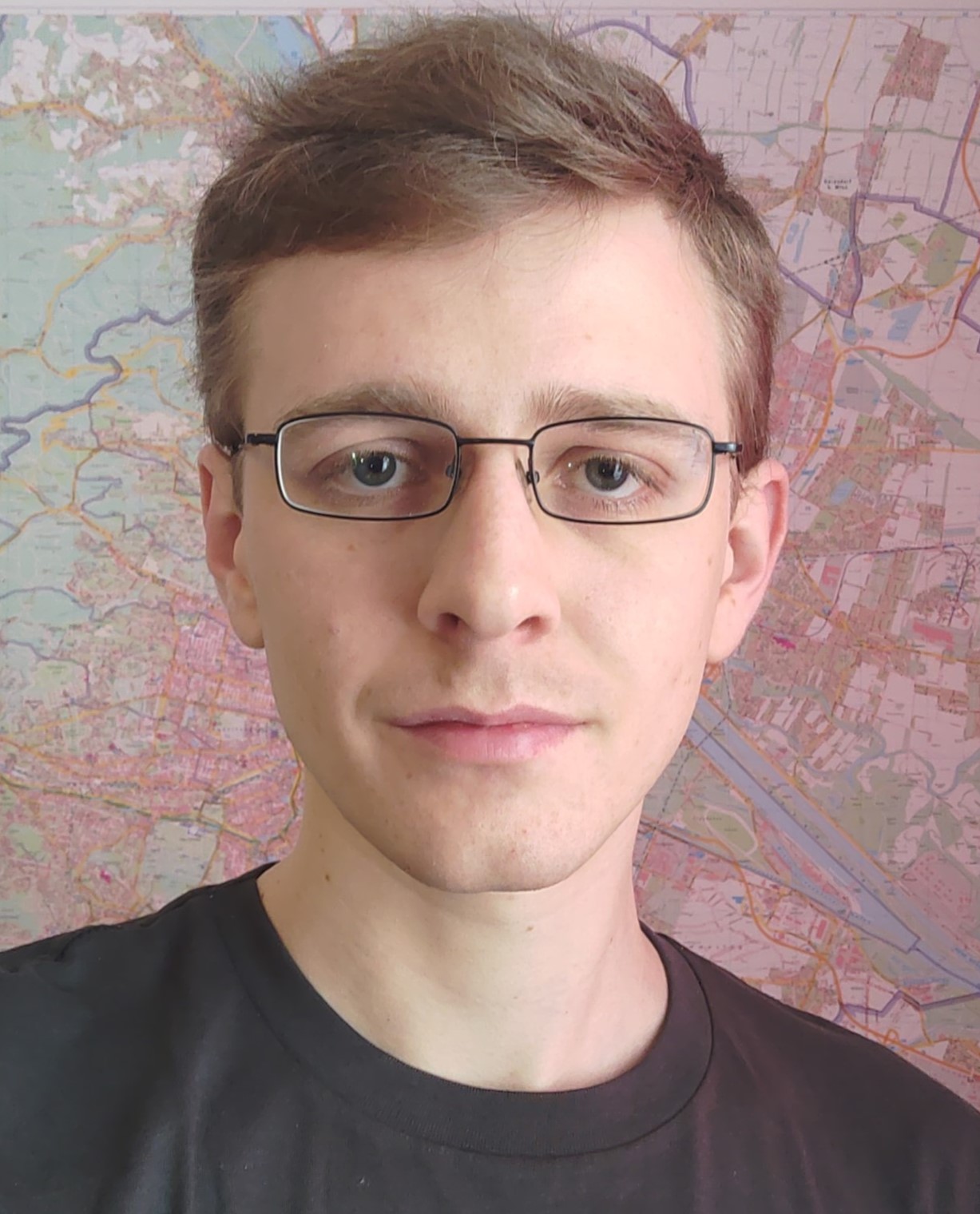}
\end{figure}
\textbf{Pirmin Urbanke} is a researcher and software engineer at Software Competence Center Hagenberg (SCCH), Austria. He received his B.Sc. and M.Sc. from the TU Wien, Austria. His current research interest is on testing distributed systems.

\begin{figure}[H]
    \includegraphics[width=0.3\textwidth]{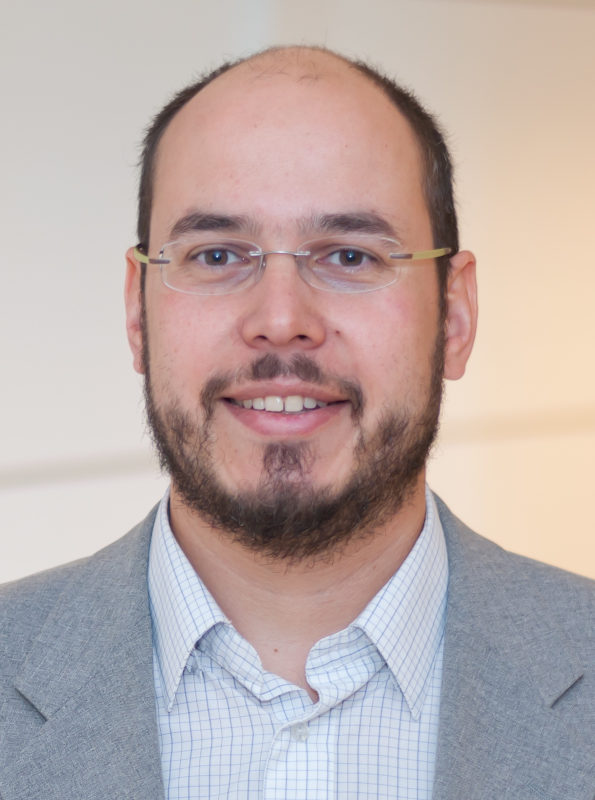}
\end{figure}
\textbf{Rudolf Ramler} is a research manager at Software Competence Center Hagenberg (SCCH), Austria. Rudolf holds a M.Sc. in Business Informatics from Johannes Kepler University Linz. He has more than 20 years of experience in applied research in the fields of software engineering, software quality assurance and testing, software analytics, and application lifecycle management. He is author of over 100 reviewed publications, co-organizer and chair of international conferences and workshops, an ISTQB certified tester, and an IEEE and ACM member. His mission and passion are to support industry in turning research results into practically successful solutions. 

\end{document}

%% file: search_strings.tex
\begin{tabularx}{\linewidth}{l X}
\toprule

Database &             Search string  \\
\midrule
Scopus & SUBJAREA (COMP) TITLE-ABS-KEY(((code) AND (test* OR model) AND (readability OR understandability OR legibility)) OR (("test" OR "code") AND (smell) AND (readab* OR understandab* OR legib*))) \\
IEEE & (("All Metadata": code) AND ("All Metadata": test* OR "All Metadata": model) AND ("All Metadata": readability OR "All Metadata": understandability OR "All Metadata": legibility)) (("All Metadata": "test" OR "All Metadata": "code") AND ("All Metadata": smell) AND ("All Metadata": readab* OR "All Metadata": understandab* OR "All Metadata": legib*)) \\
ACM & ((Title:(code) AND Title:(test* model) AND Title:(readability understandability legibility)) OR (Keyword:(code) AND Keyword:(test* model) AND Keyword:(readability understandability legibility)) OR (Abstract:(code) AND Abstract:(test* model) AND Abstract:(readability "understandability" legibility))) OR ((Abstract:("test" "code") AND Abstract:(smell) AND Abstract:(readab* understandab* legib*)) OR (Keyword:("test" "code") AND Keyword:(smell) AND Keyword:(readab* understandab* legib*)) OR (Title:("test" "code") AND Title:(smell) AND Title:(readab* understandab* legib*)))\\

\bottomrule\\
\end{tabularx}

%% file: selected_papers.tex
\tiny
\begin{tabularx}{\linewidth}{l >{\raggedright}p{4.3cm} >{\raggedright}p{1.38cm} p{0.8cm} l p{1.7cm}}
\toprule

Idx & {Title} & {Authors} & Venue & {Year} & {Study Type}   \\
\midrule 

\cite{Setiani2021251} & Developer's Perspectives on Unit Test Cases Understandability & Setiani N. et al. & ICSESS & 2021 & Experiment + Survey (hum) \\
\cite{Roy2020287} & DeepTC-Enhancer: Improving the Readability of Automatically Generated Tests & Roy D. et al. & ASE & 2020 & Experiment + Survey (hum) \\
\cite{Setiani2020169036} & Test case understandability model & Setiani N. et al. & IEEE Access & 2020 & Experiment (hum) \\
\cite{Lin2019204} & On the quality of identifiers in test code & Lin B. et al. & SCAM & 2019 & Survey (hum) \\
\cite{Alsharif2019SQL} & What Factors Make SQL Test Cases Understandable for Testers? A Human Study of Automated Test Data Generation Techniques & Alsharif A. et al. & ICSME & 2019 & Experiment + Survey (hum) \\
\cite{Leotta2018AssertJ} & Fluent vs basic assertions in Java: An empirical study & Leotta M. et al. & QUATIC & 2018 & Experiment (hum) \\
\cite{Grano2018348} & An empirical investigation on the readability of manual and generated test cases & Grano G. et al. & ICPC & 2018 & Experiment \\
\cite{Li2018Aiding} & Aiding comprehension of unit test cases and test suites with stereotype-based tagging & Li B. et al. & ICPC & 2018 & Experiment + User Study (hum) \\
\cite{Fisher2018800} & Specification-Based testing in software engineering courses & Fisher G. and Johnson C. & SIGCSE & 2018 & Experiment + Survey (hum) \\
\cite{Almasi2017263} & An industrial evaluation of unit test generation: Finding real faults in a financial application & Almasi M. et al. & ICSE-SEIP & 2017 & Experiment + Survey (hum) \\
\cite{Daka201757} & Generating unit tests with descriptive names or: Would you name your children thing1 and thing2? & Daka E. et al. & ISSTA & 2017 & Experiment + Survey (hum) \\
\cite{Bowes20179} & How Good Are My Tests? & Bowes D. et al. & WETSoM & 2017 & Concept paper (hum) \\
\cite{Palomba2016130} & Automatic test case generation: What if test code quality matters? & Palomba F. et al. & ISSTA & 2016 & Experiment \\
\cite{Li2016341} & Automatically Documenting Unit Test Cases & Li B. et al. & ICST & 2016 & User study (hum) \\
\cite{Panichella2016547} & The impact of test case summaries on bug fixing performance: An empirical investigation & Panichella S. et al. & ICSE & 2016 & Experiment (hum) \\
\cite{Zhang2016625} & Towards automatically generating descriptive names for unit tests & Zhang B. et al. & ASE & 2016 & Prototype and User Study (hum) \\
\cite{Daka2015107} & Modeling readability to improve unit tests & Daka E. et al. & ESEC/FSE  & 2015 & Experiment + Survey (hum) \\
\cite{Afshan2013352} & Evolving readable string test inputs using a natural language model to reduce human oracle cost & Afshan S. et al. & ICST & 2013 & Experiment (hum) \\
\cite{Fraser201180} & Exploiting common object usage in test case generation & Fraser G. and Zeller A. & ICST & 2011 & Experiment \\
\bottomrule
\end{tabularx}

%

%% file: factors.tex
\setlength{\tabcolsep}{0.05pt} 
\definecolor{gr}{RGB}{242,242,242} 
\rowcolors{2}{gr}{white} 
\tiny 
\renewcommand{\arraystretch}{1.5}

\begin{tabularx}{\textwidth}{Xccccccccccccccccccccc}
\toprule
\fontsize{6pt}{0cm}\selectfont
\textbf{Individual factors} & \cite{Setiani2021251} & \cite{Roy2020287} & \cite{Setiani2020169036} & \cite{Lin2019204} & \cite{Alsharif2019SQL} & \cite{Leotta2018AssertJ} & \cite{Grano2018348} & \cite{Li2018Aiding} & \cite{Fisher2018800} & \cite{Almasi2017263} & \cite{Daka201757} & \cite{Bowes20179} & \cite{Palomba2016130} & \cite{Li2016341} & \cite{Panichella2016547} & \cite{Zhang2016625} & \cite{Daka2015107} & \cite{Afshan2013352} & \cite{Fraser201180}
\\\midrule \showrowcolors
\scriptsize Test names (6) & \textbullet & \textbullet &  &  &   &   &  &   &  &  & \textbullet & \textbullet &  &  & \textbullet & \textbullet &  &  &   \\
\scriptsize Assertions (5) & \textbullet &  & \textbullet &  &  & \textbullet &  &  &  & \textbullet &  &  &  &  &  &  & \textbullet &  &  \\
\scriptsize Identifier names (5) & \textbullet & \textbullet &  & \textbullet &  &  &  &  & \textbullet &  &  & \textbullet &  &  &  &  &  &  &  \\
\scriptsize Test data (4) &  &  &  &  & \textbullet &  &  &  &  & \textbullet &  & \textbullet &  &  &  &  &  & \textbullet &  \\
\scriptsize Test summaries (4) &  & \textbullet &  &  &  &  &  & \textbullet &  &  &  &  &  & \textbullet & \textbullet &  &  &  &  \\
\scriptsize Dependencies (3) & \textbullet &  &  &  &  &  &  &  &  &  &  &  & \textbullet &  &  &  &  &  & \textbullet \\
\scriptsize Comments (2) & \textbullet &  &  &  &  &  &  &  & \textbullet &  &  &  &  &  &  &  &  &  &  \\
\scriptsize Test structure (2)& \textbullet &  &  &  &  &  &  &  &  &  &  & \textbullet &  &  &  &  &  &  &  \\

\hiderowcolors
\bottomrule\\
\end{tabularx} 

%% file: factors_models.tex
\setlength{\tabcolsep}{0.05pt} 
\definecolor{gr}{RGB}{242,242,242} 
\rowcolors{2}{gr}{white} 
\tiny 
\renewcommand{\arraystretch}{1.5}

\begin{tabularx}{\textwidth}{Xccccccccccccccccccc}
\toprule
\fontsize{6pt}{0cm}\selectfont
\textbf{Readab. models} & \cite{Setiani2021251} & \cite{Roy2020287} & \cite{Setiani2020169036} & \cite{Lin2019204} & \cite{Alsharif2019SQL} & \cite{Leotta2018AssertJ} & \cite{Grano2018348} & \cite{Li2018Aiding} & \cite{Fisher2018800} & \cite{Almasi2017263} & \cite{Daka201757} & \cite{Bowes20179} & \cite{Palomba2016130} & \cite{Li2016341} & \cite{Panichella2016547} & \cite{Zhang2016625} & \cite{Daka2015107} & \cite{Afshan2013352} & \cite{Fraser201180}   \\ \midrule \showrowcolors

\scriptsize Test  structure (3) &  &  & \textbullet &  &  &  & \textbullet &  &  &  &  &  &  &  &  &  & \textbullet &  &  \\
\scriptsize Textual features(1) &  &  &  &  &  &  & \textbullet &  &  &  &  &  &  &  &  &  &  &  &  \\

\hiderowcolors
\bottomrule\\
\end{tabularx}

%% file: grey_literature_factors_caption.tex
\caption{Factors influencing readability mapped to sources from the grey literature search. New factors identified in the gray literature are shown in purple. 
\textbf{{Str}}: structure,
\textbf{{TeN}}: test names,
\textbf{{Asse}}: assertions,
\textbf{\textcolor{purple}{Help}}: helper structures,
\textbf{{Dep}}: dependencies,
\textbf{{IdN}}: identifier names,
\textbf{\textcolor{purple}{Fix}}: fixtures,
\textbf{\textcolor{purple}{DRY}}: DRY principle,
\textbf{{TeD}}: test data,
\textbf{{Com}}: comments,
\textbf{\textcolor{purple}{DSL}}: domain specific language,
\textbf{\textcolor{purple}{Par}}: parameterized test,
{TS}: test summaries,
{TF}: textual features.}

%% file: grey_literature_factors.tex
\renewcommand{\arraystretch}{1}
\setlength{\tabcolsep}{3.4pt}
\definecolor{gr}{RGB}{242,242,242} 
\centering
\newcolumntype{Y}{>{\centering\arraybackslash}X}
\small
\rowcolors{1}{gr}{white}

\begin{tabularx}{\textwidth}{l|YYYYYYYYYYYYc}
\toprule
\hiderowcolors 
{} & \textbf{Str} & \textbf{TeN} & \textbf{Asse} & \textcolor{purple}{\textbf{Help}} & \textbf{Dep} & \textbf{IdN} & \textcolor{purple}{\textbf{Fix}} & \textcolor{purple}{\textbf{DRY}} & \textbf{TeD} & \textbf{Com} & \textcolor{purple}{\textbf{DSL}} & \textcolor{purple}{\textbf{Par}} & {TS TF} \\ \showrowcolors
 \midrule
\cite{Philip_Hauer2021} &  & \textbullet &  & \textbullet & \textbullet & \textbullet & \textbullet & \textbullet & \textbullet &   &  &  \textbullet &  \\
\cite{Fernando_Chovich_Correa2019} & \textbullet & \textbullet & \textbullet &  &  & \textbullet & \textbullet & \textbullet & \textbullet & \textbullet  &  &  &  \\
\cite{T._Yonekubo2021} & \textbullet &  &  & \textbullet & \textbullet &  & \textbullet &  & \textbullet & \textbullet  &  &  \textbullet&  \\
\cite{Arho_Huttunen2021} & \textbullet & \textbullet & \textbullet & \textbullet & \textbullet & \textbullet &  &  &  & \textbullet  &  &  &  \\
\cite{Anshul_Bansal2021} &  & \textbullet & \textbullet & \textbullet & \textbullet & \textbullet & \textbullet &  & \textbullet &   &  &  &  \\
\cite{Matheus_Rodrigues2018} & \textbullet & \textbullet &  & \textbullet &  & \textbullet & \textbullet &  & \textbullet & \textbullet  &  &  &  \\
\cite{Petri_Kainulainen2014} &  & \textbullet & \textbullet & \textbullet & \textbullet & \textbullet &  &  & \textbullet &   & \textbullet &  &  \\
\cite{Robert_C._Martin2021} & \textbullet &  & \textbullet & \textbullet & \textbullet &  & \textbullet &  &  &   & \textbullet &  &  \\
\cite{Peter_Bloomfield2020} & \textbullet & \textbullet &  &  & \textbullet & \textbullet &  &  & \textbullet &   &  &  \textbullet&  \\
\cite{Tobias_Goeschel2016} & \textbullet &  & \textbullet & \textbullet & \textbullet &  & \textbullet &  &  & \textbullet  &  &  &  \\
\cite{Urs_Enzler2014} & \textbullet & \textbullet &  &  & \textbullet & \textbullet & \textbullet &  &  &   & \textbullet &  &  \\
\cite{Maurício_Aniche_2022} &  & \textbullet &  & \textbullet & \textbullet & \textbullet &  &  & \textbullet &   &  &  &  \\
\cite{Corina_Pip2020} & \textbullet & \textbullet &  & \textbullet & \textbullet & \textbullet &  &  &  &   &  &  &  \\
\cite{Bas_Dijkstra2016} & \textbullet &  & \textbullet & \textbullet &  &  &  &  &  & \textbullet  & \textbullet &  &  \\
\cite{Henry_Coles_and_others2015} &  & \textbullet & \textbullet & \textbullet &  & \textbullet &  & \textbullet &  &   &  &  &  \\
\cite{Gil_Zilberfeld2014} & \textbullet & \textbullet & \textbullet & \textbullet &  &  & \textbullet &  &  &   &  &  &  \\
\cite{Dustin_Boswell_Trevor_Foucher2012} &  & \textbullet & \textbullet & \textbullet &  &  &  &  & \textbullet &   & \textbullet &  &  \\
\cite{Tuomas_Kareinen2012} & \textbullet &  & \textbullet &  & \textbullet &  & \textbullet & \textbullet &  &   &  &  &  \\
\cite{Vdaas_Vald_Company_blog2021} &  & \textbullet &  &  &  &  &  & \textbullet & \textbullet &   &  &  \textbullet&  \\
\cite{Carlos_Schults2021} & \textbullet & \textbullet &  &  & \textbullet &  &  &  & \textbullet &   &  &  &  \\
\cite{Jenny_Shih2020} & \textbullet & \textbullet &  &  & \textbullet &  &  &  &  & \textbullet  &  &  &  \\
\cite{Javier_Fernandes2020} & \textbullet &  &  &  &  &  & \textbullet &  &  &   & \textbullet &  \textbullet&  \\
\cite{Vadim_Bulavin2020} & \textbullet & \textbullet & \textbullet &  &  & \textbullet &  &  &  &   &  &  &  \\
\cite{Hugh_Grigg2020} & \textbullet &  & \textbullet & \textbullet &  &  &  &  &  & \textbullet  &  &  &  \\
\cite{Marc_Duiker2016} & \textbullet &  & \textbullet & \textbullet &  &  &  &  &  & \textbullet  &  &  &  \\
\cite{Lasse_Koskela2013} &  &  & \textbullet &  & \textbullet &  & \textbullet &  & \textbullet &   &  &  &  \\
\cite{Erik_Dietrich2013} & \textbullet &  &  & \textbullet &  &  & \textbullet &  &  & \textbullet  &  &  &  \\
\cite{Roberto_Casadei2013} &  &  & \textbullet &  & \textbullet &  & \textbullet &  & \textbullet &   &  &  &  \\
\cite{Dan_Carter2011} &  & \textbullet & \textbullet & \textbullet & \textbullet &  &  &  &  &   &  &  &  \\
\cite{Jason_Gorman2020} & \textbullet & \textbullet &  &  &  &  &  &  &  &   &  &  \textbullet&  \\
\cite{Anmol_Sarna2018} & \textbullet &  & \textbullet &  & \textbullet &  &  &  &  &   &  &  &  \\
\cite{Thomas_Papendieck2017} & \textbullet & \textbullet &  &  &  & \textbullet &  &  &  &   &  &  &  \\
\cite{Pawel_Lipinski2013} & \textbullet & \textbullet &  &  &  &  &  &  &  & \textbullet  &  &  &  \\
\cite{Daniel_Lindner_2013} &  &  &  & \textbullet &  & \textbullet &  &  & \textbullet &   &  &  &  \\
\cite{Stephen_Vance2013} & \textbullet &  &  &  &  & \textbullet &  &  & \textbullet &   &  &  &  \\
\cite{Mark_Needham2009} & \textbullet &  &  & \textbullet &  &  &  & \textbullet &  &   &  &  &  \\
\cite{Brooklin_Myers2021} & \textbullet & \textbullet &  &  &  &  &  &  &  &   &  &  &  \\
\cite{Daniel_Lehner2021} & \textbullet & \textbullet &  &  &  &  &  &  &  &   &  &  &  \\
\cite{Jan_Van_Ryswyck2021} &  &  &  & \textbullet &  &  &  & \textbullet &  &   &  &  &  \\
\cite{John_Ferguson_Smart_2020} &  &  & \textbullet & \textbullet &  &  &  &  &  &   &  &  &  \\
\cite{Brian_Hnat2020} &  &  &  &  &  &  &  & \textbullet &  &   &  &  \textbullet&  \\
\cite{testing_company2020} &  & \textbullet &  &  &  & \textbullet &  &  &  &   &  &  &  \\
\cite{Gleb_Bahmutov2019} &  &  & \textbullet & \textbullet &  &  &  &  &  &   &  &  &  \\
\cite{Michael_Foord2017} &  &  &  &  &  &  &  & \textbullet &  & \textbullet  &  &  &  \\
\cite{Jon_Reid2016} &  & \textbullet &  &  &  & \textbullet &  &  &  &   &  &  &  \\
\cite{Marcos_Brizeno2014} & \textbullet &  &  &  &  &  &  & \textbullet &  &   &  &  &  \\
\cite{Rafal_Borowiec2013} &  &  & \textbullet &  &  &  &  &  &  & \textbullet  &  &  &  \\
\cite{Amey_Dhoke2009} &  &  &  &  &  &  & \textbullet & \textbullet &  &   &  &  &  \\
\cite{Vladimir_Khorikov2008} &  &  &  &  &  &  &  & \textbullet &  & \textbullet  &  &  &  \\
\cite{James_Flournoy2008} &  &  &  &  &  &  & \textbullet &  &  &   &  &  \textbullet&  \\
\cite{Ev_Haus2020} &  &  &  &  &  &  &  & \textbullet &  &   &  &  &  \\
\cite{Simone_Scalabrino2019} &  &  &  &  & \textbullet &  &  &  &  &   &  &  &  \\
\cite{Jason_Roberts2019} &  &  & \textbullet &  &  &  &  &  &  &   &  &  &  \\
\cite{NAIDELE_MANJUNATH_and_OLIVIER_DE_MEULDER2019} &  &  &  &  &  &  &  &  &  &   & \textbullet &  &  \\
\cite{Derek_Snyder_and_Erik_Kuefler2019} &  &  &  &  &  &  &  & \textbullet &  &   &  &  &  \\
\cite{Adit_Lal2019} &  &  &  &  &  &  &  &  &  &   & \textbullet &  &  \\
\cite{Ryan_Cook2017} &  &  &  &  &  &  &  & \textbullet &  &   &  &  &  \\
\cite{Flavio_nickname_Company_blog2016} &  & \textbullet &  &  &  &  &  &  &  &   &  &  &  \\
\cite{Jason_Roberts2015} &  &  & \textbullet &  &  &  &  &  &  &   &  &  &  \\
\cite{Jason_Jarrett2009} &  &  & \textbullet &  &  &  &  &  &  &   &  &  &  \\
\cite{Patrick_Reagan2009} &  &  &  &  &  &  & \textbullet &  &  &   &  &  &  \\
\cite{Kristopher_Johnson2008} &  &  &  &  &  &  &  & \textbullet &  &   &  &  &  \\
 \midrule
Total & 28 & 26 & 24 & \textcolor{purple}{23} & 19 & 17 & \textcolor{purple}{17} & \textcolor{purple}{16} & 15 & 14 & \textcolor{purple}{8} & \textcolor{purple}{8} & -~~~~- \\
\bottomrule
\end{tabularx}

%% file: overlapping_factors.tex
\definecolor{gr}{RGB}{242,242,242} 

\tiny
\rowcolors{1}{gr}{white}
\begin{tabularx}{\textwidth}{XX}
    \toprule
    \hiderowcolors
    \multicolumn{1}{c}{\textbf{Scientific Literature}} & \multicolumn{1}{c}{\textbf{Grey Literature}} \\

    \midrule
    \hiderowcolors \multicolumn{2}{c}{\textbf{Test structure (5 vs 28)}} \\\showrowcolors 
    Identifier length, line length  &  \\
    Constructor calls  &  \\
    Number of identifiers  &  \\
    \cellcolor{orange!30} Control structure & \cellcolor{orange!30} Control structure \\
    \cellcolor{orange!30}Length of test case & \cellcolor{orange!30}Length of test case \\
    &  Avoid eye jumps \\
    &  Group similar test cases \\
    &  Coherent formatting \\
    &  Semantic structure (AAA, GWT, etc.) \\

    \midrule
    \hiderowcolors\multicolumn{2}{c}{\textbf{Test names (6 vs 26)}}\\\showrowcolors
    \cellcolor{orange!30}Use of patterns & \cellcolor{orange!30}Use of patterns \\
    &  Consistent naming \\
    &  Long names, spaces in names \\
    &  Include method under test in name \\
    
    \midrule
    \hiderowcolors\multicolumn{2}{c}{\textbf{Assertions (5 vs 24)}}\\\showrowcolors
    \cellcolor{orange!30}Number of assertions & \cellcolor{orange!30}One assert per test \\
    \cellcolor{orange!30}Fluent assertions & \cellcolor{orange!30}Fluent assertions \\

    \cellcolor{orange!30}Assertion messages & \cellcolor{orange!30}Assertion messages \\
    &  Appropriate assertions \\
    &  Custom assertions \\

    \midrule
    \hiderowcolors\multicolumn{2}{c}{\textbf{Helper structures (0 vs 23)}}\\\showrowcolors
    &  Builder pattern \\
    &  Composition over inheritance (of test classes) \\
    &  Methods for each step (given when then) \\
    &  Page Objects \\

    \midrule
    \hiderowcolors\multicolumn{2}{c}{\textbf{Dependencies (3 vs 19)}}\\\showrowcolors
    \cellcolor{orange!30}One test for one behavior & \cellcolor{orange!30}One test for one behavior \\

    \midrule
    \hiderowcolors\multicolumn{2}{c}{\textbf{Identifier names (5 vs 17)}}\\\showrowcolors
    \cellcolor{orange!30}Consistent & \cellcolor{orange!30}Consistent \\
    \cellcolor{orange!30}Concise & \cellcolor{orange!30}Concise \\
    \cellcolor{orange!30}Meaningful & \cellcolor{orange!30}Meaningful \\
    \cellcolor{orange!30}Use patterns & \cellcolor{orange!30}Use patterns \\

    \midrule
    \hiderowcolors\multicolumn{2}{c}{\textbf{Fixtures (0 vs 17)}}\\\showrowcolors
    &  Use setup methods, avoid fixture \\
    &  No test data in fixtures \\
    &  Avoid long fixtures \\

    \midrule
    \hiderowcolors\multicolumn{2}{c}{\textbf{DRY principle (0 vs 16)}}\\\showrowcolors
    &  Not too DRY, violate if needed \\
    &  Balance of DRY and DAMP \\

    \midrule
    \hiderowcolors\multicolumn{2}{c}{\textbf{Test data (4 vs 15}}\\\showrowcolors
    \cellcolor{orange!30}No magic values & \cellcolor{orange!30}No magic values \\
    \cellcolor{orange!30}Production like, typical, simple values & \cellcolor{orange!30}Production like, typical, simple values \\
    &  Hard coded expected values instead of computed \\
    &  Highlight important data \\

    \midrule
    \hiderowcolors\multicolumn{2}{c}{\textbf{Comments (2 vs 14)}}\\\showrowcolors
    \cellcolor{orange!30}Explanatory comments & \cellcolor{orange!30}Avoid, can become outdated \\
    & Comments for structuring (e.g., AAA pattern)  \\

    \midrule
    \hiderowcolors\multicolumn{2}{c}{\textbf{Domain specific language (0 vs 8}}\\\showrowcolors
    \cellcolor{orange!30}Explanatory comments & \cellcolor{orange!30}Avoid, can become outdated \\
    & Comments for structuring (e.g., AAA pattern)  \\

    \midrule
    \hiderowcolors\multicolumn{2}{c}{\textbf{Parameterized test (0 vs 8}}\\\showrowcolors
    & Data driven tests to reduce code duplication \\

    \midrule
    \hiderowcolors\multicolumn{2}{c}{\textbf{Test summaries (4 vs 0)}}\\\showrowcolors
    Use source code summarization techniques & \\

    \midrule
    \hiderowcolors\multicolumn{2}{c}{\textbf{Textual features (1 vs 0)}}\\\showrowcolors
    Natural language processing, dictionaries & \\

    \bottomrule

\end{tabularx}

%% file: test_cases_origin_factors.tex
\renewcommand{\arraystretch}{1.1}
\definecolor{gr}{RGB}{242,242,242} 
\centering
\tiny
\rowcolors{2}{gr}{white}

\begin{tabular}{p{1.3cm}p{2.6cm}p{2cm}p{4.3cm}}
\toprule 
\textbf{Influence Factor} & \textbf{Test Name} & \textbf{Origin Project} & \textbf{Modification A/B}\\
\midrule
Structure & testPrimitiveTypeClass Serialization  & Apache Commons Lang3 & Loops vs. unrolled loops \\
Structure & testReducedFraction & Apache Commons Lang3 &  Loops vs. unrolled loops (exemplary values) \\
Structure & testContainsIgnoreCase \_LocaleIndependence  & Apache Commons Lang3 &  Loops vs. unrolled loops \\
Assertions & test10  & Apache Commons Lang3 & Try catch vs. AssertThrows \\
Assertions & test2 & Apache Commons Lang3 & Try catch vs. AssertThrows \\
Assertions & test303 & Apache Commons Lang3 & Try catch vs. AssertThrows \\
Structure & testInvert & Apache Commons Lang3 & Variable reuse \\
Structure & testNegate & Apache Commons Lang3 & Variable reuse \\
Structure & testAbs & Apache Commons Lang3 & Variable reuse \\
Structure & test551 & Apache Commons Lang3 & Remove package names, if-structure, system out print and helper variables \\
Structure & test0074 & Apache Commons Lang3 & Remove package names, if-structure, system out print and helper variables \\
Structure & test1113 & Apache Commons Lang3 & Remove package names, if-structure, system out print and helper variables \\
Comment & testContainsRange  & Apache Commons Lang3 & Remove comments \\
Comment & testFactory\_double & Apache Commons Lang3 & Remove comments \\
Comment & testWrap\_StringInt StringBooleanString & Apache Commons Lang3 & Remove comments \\ 
Parameterized & testPrimitiveTypeClass Serialization  & Apache Commons Lang3 & Loops vs. Parameterized. Replace with @MethodSource \\
Parameterized & testReducedFraction & Apache Commons Lang3 & Loops vs. Parameterized. Replace with @MethodSource (chained stream) \\
Parameterized & testContainsIgnoreCase \_LocaleIndependence  & Apache Commons Lang3 & Loops vs. Parameterized. Replace with inlined CSV \\
Dependencies & testAllNullBooleans & Apache Flink & Split up tests \\
Dependencies & testSerializeAndParse & Protoclbuffers Protobuf & Split up tests (original has comments) \\
Dependencies & testSetContentObject & Apache Commons Email & Split up tests (original has comments) \\
Assertions & testFourElement2 & JetBrains IntelliJ Community & Specific assertions (JUnit vs. Hamcrest/AssertJ)\\
Assertions & showsAllStsGaDownloads & Dchartfield Sagan & Specific assertions (JUnit vs. Hamcrest/AssertJ) \\
Assertions & indexedReadAndIndexed WriteMethods & Spring Framework & Specific assertions (JUnit vs. Hamcrest/AssertJ) \\
Structure & testChoicesWithValid DefaultValue & Apache Flink & Remove unnecessary try catch \\
Structure & testApplyToMovesValue PassedOnShortName ToLongNameIfLong NameIsUndefined & Apache Flink & Remove unnecessary try catch \\
Structure & testApplyToWithMultiple Types & Apache Flink & Remove unnecessary try catch \\
Fixture, test data & Student Example 03 & Student Solution & Remove fixture and member variables or constants \\
Fixture, test data & Student Example 02 & Student Solution & Remove fixture and member variables or constants or constants \\
Fixture, test data & Student Example 01 & Student Solution & Remove fixture and member variables or constants \\
\bottomrule
\end{tabular}

%% file: general_software_development_exp_years.tex
\begin{tabular}{lrr}
\toprule
\textbf{Years} &  \textbf{Absolute} &  \textbf{Percentage} \\
\midrule
    0 &         0 &          0\% \\
  1-2 &         2 &        2.6\% \\
  2-5 &        41 &       53.2\% \\
   \textgreater{}5 &        34 &       44.2\% \\\midrule
\textbf{ Sum:} &        77 &      100.0\% \\
\bottomrule
\end{tabular}

%% file: professional_software_development_exp_years.tex
\begin{tabular}{lrr}
\toprule
\textbf{Years} &  \textbf{Absolute} &  \textbf{Percentage} \\
\midrule
  0 &        20 &       26.0\% \\
  1-2 &        24 &       31.2\% \\
  2-5 &        25 &       32.5\% \\
  \textgreater{}5 &         8 &       10.3\% \\\midrule
\textbf{ Sum:} &        77 &      100.0\% \\
\bottomrule
\end{tabular}

%% file: stat_analysis_likert.tex
\setlength{\tabcolsep}{2.5pt} 
\definecolor{gr}{RGB}{242,242,242} 
\rowcolors{2}{gr}{white} 
\footnotesize 

\begin{tabularx}{\textwidth}{Xrrrrrrrr}
    \toprule
    \hiderowcolors
    & \multicolumn{3}{c}{\small \textbf{A}} & \multicolumn{3}{c}{\small\textbf{B}} & \multicolumn{2}{c}{\small\textbf{Compare}} \\
    \small Modification A/B (Influence Factor) &\small N &\small med &\small sd & \small N &\small med &\small sd &\small $p$ &\small $\delta$ \\ \midrule \showrowcolors
\small\textbf{Loops vs. Unrolled Loops} (Structure) & 29 & 3 & 1.2 & 29 & 4 & 1.1 & \textbf{0.02} & \textbf{-0.35} \\
testPrimitiveTypeClassSerialization & 9 & 4 & 0.8 & 11 & 4 & 0.9 & 0.21 &  \\
testReducedFraction & 9 & 3 & 0.9 & 9 & 3 & 1.3 & 0.89 &  \\
testContainsIgnoreCase\_LocaleIndependence & 11 & 2 & 1.2 & 9 & 4 & 0.9 & \textbf{0.01} & \textbf{-0.67} \\

\midrule\small\textbf{Try Catch vs. AssertThrows} (Assertions)& 27 & 2 & 0.9 & 28 & 2 & 1.1 & 0.31 &  \\
test10 & 9 & 2 & 0.9 & 10 & 2 & 1.2 & 0.90 &  \\
test2 & 9 & 2 & 0.8 & 9 & 3 & 1.1 & \textbf{0.04} & \textbf{-0.54} \\
test303 & 9 & 2 & 1 & 9 & 2 & 0.9 & 0.89 &  \\

\midrule\small\textbf{Variable Reuse} (Structure) & 28 & 4 & 0.8 & 29 & 4 & 0.8 & 0.32 &  \\
testInvert & 9 & 4 & 0.7 & 11 & 3 & 0.9 & 0.54 &  \\
testNegate & 9 & 4 & 0.9 & 9 & 4 & 0.5 & 0.15 &  \\
testAbs & 10 & 4 & 0.7 & 9 & 4 & 0.9 & 0.78 &  \\

\midrule\small\textbf{Package Names, If-Structure,..} (Structure) & 28 & 1 & 0.7 & 28 & 2 & 0.9 & \textbf{0.00} & \textbf{-0.59} \\
test551 & 8 & 1 & 0.7 & 11 & 3 & 0.8 & \textbf{0.00} & \textbf{-0.82} \\
test0074 & 9 & 1 & 0.9 & 8 & 2 & 0.8 & 0.16 &  \\
test1113 & 11 & 1 & 0.5 & 9 & 2 & 0.4 & \textbf{0.03} & \textbf{-0.51} \\

\midrule\small\textbf{Remove Comments} (Comments)& 28 & 4 & 1.1 & 28 & 3 & 0.8 & \textbf{0.02} & \textbf{0.36} \\
testContainsRange & 9 & 4 & 0.9 & 10 & 4 & 0.7 & 0.43 &  \\
testFactory\_double & 9 & 5 & 1.1 & 9 & 3 & 0.7 & 0.08 &  \\
testWrap\_StringIntStringBooleanString & 10 & 4 & 1.4 & 9 & 2 & 0.9 & 0.15 &  \\

\midrule\small\textbf{Loops vs. Parameterized} (Parameterized)& 45 & 2 & 1.4 & 44 & 4 & 1.2 & \textbf{0.00} & \textbf{-0.34} \\
testPrimitiveTypeClassSerialization & 14 & 4 & 1.1 & 15 & 4 & 0.9 & 0.85 &  \\
testReducedFraction & 14 & 2 & 1.4 & 14 & 2 & 1.2 & 0.77 &  \\
testContainsIgnoreCase\_LocaleIndependence & 17 & 1 & 0.8 & 15 & 4 & 1.1 & \textbf{0.00} & \textbf{-0.84} \\

\midrule\small\textbf{Split Up Tests} (Dependencies) & 45 & 4 & 1.3 & 48 & 4 & 0.8 & \textbf{0.00} & \textbf{-0.33} \\
testAllNullBooleans & 14 & 4 & 1.4 & 17 & 4 & 0.7 & 0.08 &  \\
testSerializeAndParse & 15 & 4 & 1.3 & 16 & 4 & 0.9 & 0.60 &  \\
testSetContentObject & 16 & 3 & 1.2 & 15 & 4 & 0.8 & \textbf{0.01} & \textbf{-0.50} \\

\midrule\small\textbf{Specific Assertion} (Assertions) & 44 & 3 & 1.2 & 48 & 2 & 1.1 & 0.93 &  \\
testFourElement2 & 16 & 3 & 1 & 17 & 2 & 1.2 & 0.44 &  \\
showsAllStsGaDownloads & 14 & 3 & 1.2 & 16 & 3 & 0.8 & 0.73 &  \\
indexedReadAndIndexedWriteMethods & 14 & 2 & 1.3 & 15 & 3 & 1.1 & 0.47 &  \\

\midrule\small\textbf{Unnecessary Try Catch} (Structure)& 47 & 3 & 1.1 & 48 & 3 & 1.3 & 0.12 &  \\
testChoicesWithValidDefaultValue & 16 & 4 & 0.9 & 17 & 4 & 1.1 & 0.90 &  \\
testApplyToMovesValuePassedOnShortName... & 15 & 2 & 1.1 & 16 & 2 & 1.1 & 0.90 &  \\
testApplyToWithMultipleTypes & 16 & 2.5 & 1.1 & 15 & 4 & 1.2 & \textbf{0.01} & \textbf{-0.53} \\

\midrule\small\textbf{Remove Fixture} (Fixture, Test Data)& 46 & 4 & 1.1 & 46 & 4 & 0.9 & 0.87 &  \\
Student Example 03 & 16 & 4 & 1.3 & 16 & 4 & 1.1 & 0.54 &  \\
Student Example 02 & 15 & 4 & 0.9 & 15 & 5 & 0.7 & 0.17 &  \\
Student Example 01 & 15 & 5 & 0.9 & 15 & 4 & 0.8 & 0.10 &  \\

    \bottomrule
\end{tabularx}